\documentclass[11pt,3p,review,authoryear]{elsarticle}

\usepackage{amsmath}
\usepackage[singlelinecheck=false]{caption}
\usepackage{multirow}
\usepackage{adjustbox}
\usepackage{pdflscape}
\usepackage[colorlinks,allcolors=red]{hyperref}
\usepackage{hhline}
\usepackage{mathtools}

\DeclarePairedDelimiter\floor{\lfloor}{\rfloor}
\usepackage{titlesec}
\titleformat{\paragraph}
{\normalfont\normalsize\bfseries}{\theparagraph}{1em}{}
\titlespacing*{\paragraph}
{0pt}{3.25ex plus 1ex minus .2ex}{1.5ex plus .2ex}

\begin{document}

\begin{frontmatter}

\title{Inside the black box: Neural network-based real-time prediction of US recessions}

\author[inst1]{Seulki Chung}

\affiliation[inst1]{organization={GSEFM, Department of Empirical Economics, Technische Universität Darmstadt},
            addressline={Karolinenpl.5}, 
            city={Darmstadt},
            postcode={64289}, 
            country={Germany}}



\begin{abstract}
Long short-term memory (LSTM) and gated recurrent unit (GRU) are used to model US recessions from 1967 to 2021. Their predictive performances are compared to those of the traditional linear models. The out-of-sample performance suggests the application of LSTM and GRU in recession forecasting, especially for longer-term forecasts. The Shapley additive explanations (SHAP) method is applied to both groups of models. The SHAP-based different weight assignments imply the capability of these types of neural networks to capture the business cycle asymmetries and nonlinearities. The SHAP method delivers key recession indicators, such as the S\&P 500 index for short-term forecasting up to 3 months and the term spread for longer-term forecasting up to 12 months. These findings are robust against other interpretation methods, such as the local interpretable model-agnostic explanations (LIME) and the marginal effects.  
\linebreak
\end{abstract}


\end{frontmatter}
\newpage
\section{Introduction}

Recession forecasting has been a longstanding challenge for policymakers and market practitioners, enabling them to make timely decisions that could mitigate the impact of a recession. However, due to the interconnected nature of modern economics, this task has proven difficult and has had limited success. Treating recession as a binary event, traditional linear methods such as probit and logit models and extensions have been widely employed for this forecasting task. However, in recent decades, machine learning techniques, including artificial neural networks, have become increasingly popular among economists and have been applied to certain macroeconomic forecasting problems. Nevertheless, whether and why these modern approaches allow for informed deviation from conventional linear models when predicting recessions has yet to be demonstrated.

Since the pioneering work of \citet{mitchell1938}, which identified 21 variables out of a larger set as potential economic indicators for business cycles, researchers have been engaged in selecting indicators and developing theoretical frameworks and predictive models to link these indicators with business cycles. However, most studies have primarily focused on linear frameworks. Many have used probit regression models and extensions to generate recession forecasts. \citet{estrella1996, estrella1998} compiled a combination of financial and macroeconomic variables and conducted recession forecasting using a probit framework. Their findings revealed that stock prices exhibit greater short-term predictability, while the yield curve slope performs better for longer-term predictions. \citet{wright2006} demonstrated that probit models incorporating the federal funds rate and term spread as predictors outperform models solely relying on the term spread. \citet{dueker1997, dueker2002} expanded the standard probit model by incorporating Markov regime switching within the probit framework, allowing for coefficient variation. \citet{chauvet2005} introduced several specifications to the probit model that accounted for different business cycle dependencies and autocorrelation errors, concluding that their more complicated extensions improved the accuracy of recession forecasts. \citet{fornari2010} and \citet{nyberg2014} incorporated vector autoregression components into the probit model to capture the endogenous dynamics of the predictors.

While probit regression and its extensions are widely used in business cycle forecasting, there has been a growing interest in exploring the predictability of nonlinear models, including machine learning methods. This interest stems from recognizing that the business cycle often exhibits asymmetric and nonlinear patterns (\citet{acemoglu1997, morley2012}). Empirical evidence supports this notion, such as the work of \citet{tiao1994}, who demonstrate that a threshold autoregressive model outperforms a linear autoregressive model for predicting GDP growth. \citet{maasoumi1994} examine multiple macroeconomic time series and confirm their nonlinear nature. \citet{puglia2021} highlight the attractiveness of machine learning methods as an alternative to probit regression and its extensions, noting that probit methods typically require additional parameters for flexibility. In contrast, flexibility is inherent in machine learning methods. \citet{stock1998} compare the forecasting performance of 49 univariate linear and nonlinear models across 215 macroeconomic time series. They find that some nonlinear models perform poorly compared to linear models. \citet{jaditz1998} explore using nearest neighbor regression models for forecasting industrial production but observe only marginal improvements in predictive performance. \citet{vishwanathan2002} present an iterative algorithm for support vector machines in classification problems. \citet{ng2014} applies a tree ensemble classifier to a large panel of predictors. \citet{fornaro2016} combines a Bayesian methodology with a shrinkage prior within the probit framework to predict recessions using extensive predictors. More recently, \citet{holopainen2017}, \citet{bluwstein2020}, and \citet{vrontos2021} employ various machine learning methods for economic event forecasting. \citet{vrontos2021} provide empirical evidence supporting the application of machine learning over traditional econometric techniques in recession forecasting.

As a subfield of machine learning, neural networks have gained significant attention and application in fields like finance, primarily due to high pattern recognition capabilities and their ability to establish flexible mappings between the variables (\citet{zhang1998}). Neural networks, being highly nonlinear and nonparametric, can approximate almost any functional form accurately, as stated by the universal approximation theorem (\citet{hornik1989}), given that the network is wide or deep enough. Consequently, they are valuable modeling tools, particularly when there is limited prior knowledge about the appropriate functional relationships. However, using neural networks in macroeconomic studies has been relatively limited due to the small sample sizes and low-frequency nature of macroeconomic data. \citet{swanson1997} compare artificial neural networks with linear models regarding predictive performance for nine macroeconomic variables, revealing only marginal improvements in forecast accuracy. \citet{moshiri2000} apply neural networks to inflation forecasting using a dataset of 300 observations spanning 25 years of monthly data. \citet{tkacz2001} compares multivariate neural networks with linear and univariate models, finding minor forecast improvements in the short term but more pronounced benefits for longer horizons, such as a one-year forecast. \citet{qi2001} employs a simple feed-forward neural network to predict US recessions using a range of financial and economic indicators, identifying some indicators as useful for prediction. More recently, \citet{puglia2021} compare neural networks with probit regression in forecasting US recessions using the term spread and other macro-financial variables, finding little difference between the models when evaluated. \citet{wang2022} employ a specific type of recurrent neural network, namely a Bi-LSTM with autoencoder, along with other machine learning models to predict the beginning and end of economic recessions in the US. Their results suggest that the Bi-LSTM with autoencoder is the most accurate model. 

The novelty of this paper is twofold: Firstly, it focuses on two special types of recurrent neural networks, the long short-term memory (LSTM) and the gated recurrent unit (GRU), which address the limitations of a standard recurrent neural network related to the exploding and vanishing gradient problems. Their performance is compared to the simple feedforward neural network (FFN), which suffers from the key limitation of specifying the temporal dependence upfront in the model's design and to the traditional linear models in the context of recession forecasting. Secondly, the paper applies the Shapley additive explanations method (SHAP) to GRU, which shows higher overall performance than LSTM, to explore the variable importance of different forecast horizons. In recession forecasting, \citet{puglia2021} and \citet{delgado2022} also use the SHAP method to decompose recession forecasts, but they are applied to models other than LSTM and GRU. The three main findings can be summarized as follows: Firstly, the out-of-sample performance strongly supports the application of LSTM and GRU in recession forecasting, especially for long-term forecasting tasks. They outperform other types of models across five forecast horizons for different types of statistical performance metrics. Secondly, GRU and the ridge logit model differ in assessing variable importance, evident in the different variable orders based on the SHAP values. Lastly, while the leading predictors for GRU and ridge logit models slightly differ, key indicators like S\&P 500 index, real GDP, and private residential fixed investment consistently emerge for short-term predictions (up to 3 months). The term spread and producer price index precede in longer-term forecasts (6 months or more). These results are corroborated by local interpretable model-agnostic explanations (LIME) and marginal effects.

The remainder of the paper is as follows. Section 2 explains the data used. Section 3 describes the models and performance evaluation metrics and outlines the research methodology. Section 4 presents the prediction results. Section 5 reports the methodology and results of SHAP and other interpretation methods, and Section 6 concludes.

\section{Data}

Prior studies in business cycle forecasting often rely on macroeconomic indicators, subject to revisions after initial estimates. \citet{stark2002} demonstrate that the accuracy of forecasts is influenced by using the most up-to-date data instead of real-time data. Therefore, when comparing forecasts from new models to benchmark forecasts, it is crucial to ensure that the comparisons are based on real-time data. My research focuses primarily on assessing the real-time predictability of neural network models for the Great Recession and the COVID-19 recession in the United States. This necessitates working with real-time data since the information available in hindsight was inaccessible before the recession. To evaluate the predictability of the recessions, I utilized the same data available to real-time forecasters for out-of-sample forecasting.

The dataset employed for prediction consists of 194 real-time vintages of macroeconomic and financial market variables, covering the period from February 1967 to October 2021. The out-of-sample period begins in November 2006. Considering previous studies and real-time data availability, a set of 25 predictors is chosen. A detailed list with descriptions of these variables can be found in Table 1.

\begin{table}[!ht]
    \caption{Overview of predictors}
    \resizebox{\textwidth}{!}{\begin{tabular}{llllll}
    \hline
    Nr. & Predictive variable & Abbreviation & Category & Transformation & Frequency \\
    \hline
    1 & Average hourly earnings of production and nonsupervisory employees & AHETPI & Income & Percent change & Monthly\\
    2 & Average weekly hours of production and nonsupervisory employees & AWHNONAG & Labor market & percent change & Monthly \\
    3 & Moody's BAA yield & BAA &Money and credit & First-order difference & Monthly\\
    4 & Moody's BAA yield relative to 10-Year treasury yield & BAA10YM & Money and credit & First-order difference & Monthly \\
    5 & Real manufacturing and trade industries sales & CMRMTSPL & Output & Log growth rate & Monthly \\
    6 & Corporate profits after tax & CP & Income & Log growth rate & Quarterly \\ 
    7 & Real disposable personal income & DSPIC96 & Income &Log growth rate & Monthly \\
    8 & Effective federal funds rate & FEDFUNDS & Financial market & First-order difference & Monthly \\
    9 & Real gross domestic product & GDPC1 & Output & Log growth rate & Quarterly \\
    10 & Privately-owned housing units started & HOUST & Housing market & Log growth rate & Monthly \\
    11 & Industrial production index & INDPRO & Output & Log growth rate & Monthly \\
    12 & Real M1 money stock & M1REAL & Money and credit & First-order difference & Monthly\\
    13 & Real M2 money stock & M2REAL & Money and credit & First-order difference & Monthly\\
    14 & Non-farm payroll total & PAYEMS & Labor market & Log growth rate & Monthly\\
    15 & Real personal consumption expenditures & PCEC96 & Prices & Log growth rate & Monthly \\
    16 & Privately-owned housing units permitted & PERMIT & Housing market & Log growth rate & Monthly \\
    17 & Producer price index by all commodities & PPIACO & Prices & Log growth rate & Monthly \\
    18 & Private residential fixed investment & PRFI &Housing market & Log growth rate & Quarterly \\
    19 & S\&P 500 index & SP500 & Financial market & Log growth rate & Daily \\
    20 & 3-month treasury bill rate  & TB3MS &Financial market & First-order difference & Monthly \\
    21 & Term spread - 5-year treasury yield minus 3-month treasury bill rate & T5Y3MM & Financial market & First-order difference & Monthly \\
    22 & Consumer Sentiment - University of Michigan & UMCSENT & Prices & Log growth rate & Monthly \\
    23 & Unemployment rate & UNRATE & Labor market & First-order difference & Monthly \\
    24 & Producer price index by commodity: final demand: finished goods & WPSFD49207 & Prices & Log growth rate & Monthly \\
    25 & Real personal income excluding current transfer receipts & W875RX1 & Income & Log growth rate & Monthly\\
    \hline
        \vspace{-1cm}
    \end{tabular}}
    \caption*{The table presents a list of predictive variables in alphabetical order based on their abbreviations according to ALFRED. It includes information about their respective categories, transformations applied to ensure stationarity, and their original data frequency.}
\end{table}

The selected predictors cover various categories: output, income, prices, labor, housing, money and credit, and financial markets. The data frequency varies from daily to quarterly. Higher frequency data is aggregated into monthly data using the mean. Quarterly frequency variables are transformed into monthly equivalents using natural cubic spline interpolation. Specifically, at each month, all available data up to that point are used to calculate the interpolating cubic spline. This spline curve is then utilized to generate data of monthly frequency that lie on the curve between the quarterly data points.

Most monthly data vintages for the variables are obtained from Archival Federal Reserve Economic Data (ALFRED). However, there are two groups of variables for which the real-time data before 2013 is not available in ALFRED. The first set of variables, including real personal income excluding transfer receipts, real manufacturing and trade sales, total non-farm payroll employment, and the industrial production index, are used by \citet{chauvet2008} to identify business cycle dates in real time. The real-time data for these series, provided by Jeremy Piger on his website, ends in August 2013. However, it can be easily extended beyond 2013 using the ALFRED data. The second set of variables consists of real M1 and M2 money stock, for which the earliest available real-time data in ALFRED is from January 2014. To address the absence of real-time vintages for real M1 and M2 money stocks before this date, nominal M1 and M2 money stocks are adjusted for inflation using the consumer price index, which has real-time vintages readily accessible in ALFRED.

To identify recession periods in the United States, I rely on the business cycle expansion and contraction dates determined by the National Bureau of Economic Research (NBER). NBER is the standard reference for US business cycles in the existing literature. This study defines recession months as the period following the peak and continuing until the trough, while all other months are considered periods of economic expansion. The earliest available vintage of the NBER recession indicator in ALFRED is September 2014. For the monthly vintages preceding that date, I manually collect and construct them based on the official announcements made by the NBER business cycle dating committee\footnote{\url{https://www.nber.org/research/business-cycle-dating/business-cycle-dating-committee-announcements}}. 

\begin{table}[!ht]
        \caption{US Business Cycle dates} 
    \begin{tabular}{llll}
    \hline
    Date & Type & Duration & Announcement \\
    \hline
    1980:01 & Peak & 6 & 1980:06(+5) \\
    1980:07 & Trough & 12 & 1981:07(+12) \\
    1981:07 & Peak & 16 & 1982:01(+6) \\
    1982:11 & Trough & 92 & 1983:07(+8) \\
    1990:07 & Peak & 8 & 1991:04(+9)\\
    1991:03 & Trough & 120 & 1992:12(+21)\\
    2001:03 & Peak & 8 & 2001:11(+8)\\
    2001:11 & Trough & 73 & 2003:07(+20)\\
    2007:12 & Peak & 18 & 2008:12(+12)\\
    2009:06 & Trough & 128 & 2010:09(+15)\\
    2020:02 & Peak & 2 & 2020:06(+4)\\
    2020:04 & Trough & ongoing & 2021:07(+15)\\
    \hline
    \vspace{-1cm}
    \end{tabular} \\
    \caption*{The table reports NBER business cycle dates, including the type of cycle (contraction or expansion), the duration in months, and the time of announcement. The data covers the period from 1980 to 2021; publication lags in months are indicated in parentheses.}
\end{table} 

One major practical challenge with NBER business cycle dates is that they are often announced with significant publication delays. Table 2 presents the peak and trough dates of the US business cycle from 1980 to 2021, along with their corresponding announcement dates. The publication lags for the recessions in the table range from 5 to 21 months, with troughs being identified later than peaks on average. While the NBER business cycle dates remain unchanged once finalized, publication lags complicate the creation of real-time versions of the NBER recession indicator. To address this, the NBER recession indicator in ALFRED is constructed assuming the previous state remains unchanged until a new turning point is officially announced.

\section{Econometric methodology}

This section focuses on the technical aspects of the models used to predict the two most recent recessions in the United States. The paper explores neural networks, complex nonlinear models of interconnected nodes arranged in multiple layers. Given that the network is wide or deep enough, these networks can approximate any linear or nonlinear continuous functions, as stated by the universal approximation theorem (\citet{hornik1989}). In this paper, three types of neural networks are employed for recession forecasting: feedforward neural network (FFN), long short-term memory (LSTM), and gated recurrent unit (GRU). Section 3.1 provides a detailed description of these models. Model specifications, including estimation and prediction techniques, are discussed in Section 3.2. Finally, Section 3.3 introduces statistical measures that effectively evaluate the prediction performance.

\subsection{Neural Networks}

\subsubsection{Feedforward neural network}  
\begin{figure}[!ht]
\caption{The architecture of a feed-forward neural network}
\includegraphics[width=\textwidth]{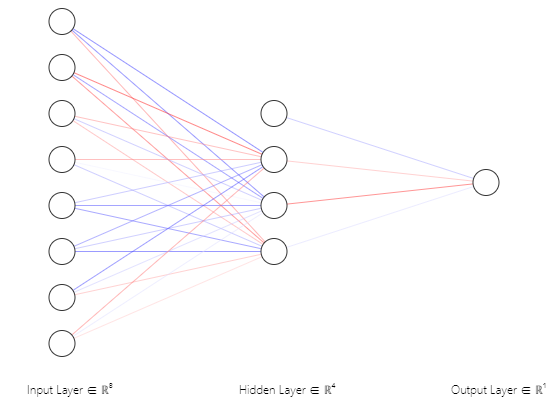}\\
\caption*{The figure illustrates an example of a feed-forward neural network featuring one hidden layer and bias units. The output layer consists of a single unit that uses a sigmoid function for activation, catering to a binary classification task. For the sake of illustration, the edges are depicted in different colors to indicate diverse edge weights, which can be either positive or negative, and also in different opacities to highlight varying magnitudes of the edge weights.}
\end{figure}

FFN is a widely used and straightforward type of artificial neural network. It consists of multiple processing units called nodes or neurons organized into layers. It operates by transmitting information in a unidirectional manner, where data flows from the input layer to the output layer without feedback loops. Figure 1 illustrates, based on the NN-SVG \footnote{\citet{lenail2019} \url{https://doi.org/10.21105/joss.00747}}, an example of a three-layer FFN designed for binary classification. It comprises an input layer with eight units, a hidden layer with four units, and an output layer with one unit. The first units in the input and hidden layers serve as bias units. In this configuration, data from the input layer is passed through the hidden layer, which transforms the data. The values obtained from the hidden layer are then forwarded to the output layer, translating into desired outputs based on the problem. In the case of binary classification, the last unit in the output layer utilizes a sigmoid activation function, producing a value between 0 and 1. This value represents the probability of an event occurring.

Given this architecture, the unknown underlying function $f$ for an output node can be written as 
\begin{align*}
    f(X)=g_2\Bigg[\alpha_0+\sum_{j=1}^k\alpha_j g_1\Bigg(\beta_{0j}+\sum_{i=1}^n\beta_{ij} x_i\Bigg)\Bigg]+\epsilon,
\end{align*}
where $n$ is the number of predictors, $k$ is the number of units in the hidden layer, $g_1$ is the activation function in the hidden layer, $g_2$ is the activation function in the output layer, $\beta_{ij}$ and $\alpha_{j}$ represent the weight parameters from the input to the hidden layer and from the hidden to the output layer, respectively, and $\epsilon$ is the error term. The backpropagation process estimates the weight parameters, which seeks to repeatedly update the weights until convergence based on the derivatives of the cost function for input and the hidden layers.

\subsubsection{Long short-term memory}

FFN is restricted to one-way signal flow, meaning there is no feedback mechanism where the output of a layer can influence the same layer. Consequently, FFN cannot capture temporal dependencies in time series data. Although time series data can be fed into an FFN by incorporating additional input units representing previous time points, the main limitation lies in the fixed dimensionality of inputs and outputs. In other words, the precise length of temporal dependence must be predetermined, which is often unknown in real-world scenarios. This is where RNN comes into play. RNN establishes connections that form cycles, allowing for feedback loops where data can be fed back into the input before being forwarded again. This feedback loop enables RNN to maintain an internal state or memory to process sequences of inputs or time steps. Theoretically, RNN can retain all information over time and handle long-term dependencies. However, they face two computational challenges. Firstly, as input sequences grow longer, the backpropagation process relies heavily on the chain rule, which may lead to vanishing gradients. If any gradient approaches zero, all other gradients will diminish exponentially fast due to the multiplicative nature of the chain rule. This phenomenon, known as the vanishing gradient problem, prevents effective training in the model. Secondly, depending on the length of input sequences, the gradient of the loss function can become excessively large and result in numerical instability, referred to as the exploding gradient problem.

\begin{figure}[!ht]
\caption{The architecture of a long short-term memory network}
\includegraphics[width=\textwidth]{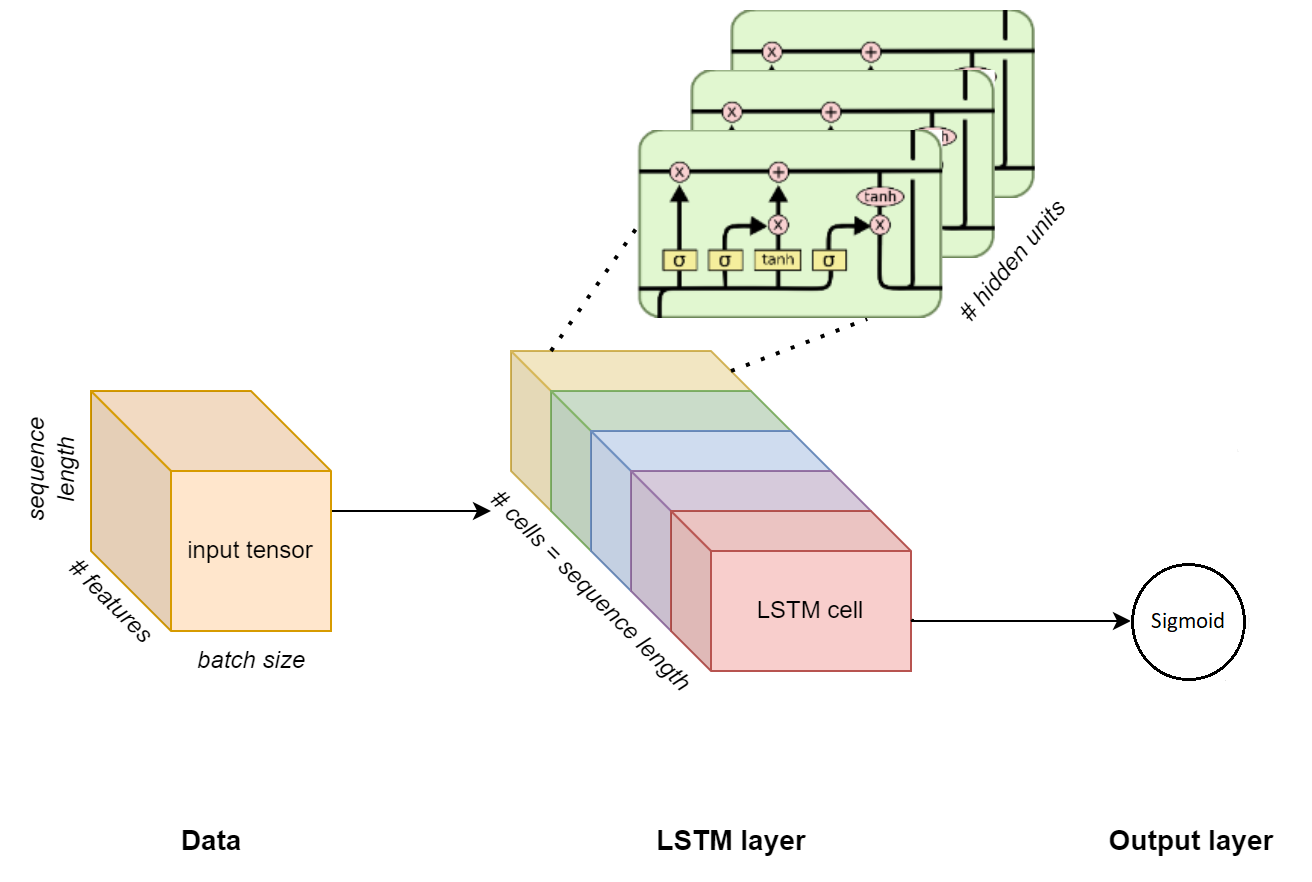}\\
\caption*{The figure depicts a basic LSTM neural network architecture comprising an LSTM layer and an output layer. The data is represented as a three-dimensional tensor, with dimensions for batch size, variables, and time steps. Each LSTM cell in the LSTM layer handles information retrieval at a specific time point using specialized processing gates in the hidden units.}
\end{figure}

\citet{hochreiter1997} introduced the long short-term memory (LSTM) architecture and a corresponding learning algorithm to address the challenges of error back-flow and long-term dependency in RNN. Figure 2 illustrates a basic LSTM neural network comprising an LSTM and output layer. The input data is represented by a three-dimensional tensor, with dimensions for batch size, variables, and time steps. The LSTM layer consists of as many LSTM cells as there are time steps, with each cell responsible for information retrieval at a specific time point in a time series. These cells contain hidden units comprising special nodes and gates designed to process the information. Figure 3 provides a closer look at the hidden units of LSTM and GRU. For both Figures 2 and 3, the graphics of the hidden units are adapted from Chris Colah's blog\footnote{\url{https://colah.github.io/posts/2015-08-Understanding-LSTMs/}}. The LSTM unit on the left features three gates (forget, input, and output gates) denoted by red dotted lines and a cell state, a crucial distinction between RNN and LSTM networks. The cell state functions like a conveyor belt that extends across the entire sequence, facilitating the retention of information over long periods. This mechanism closely resembles the long-term memory function of the human brain. The three gates regulate the flow of information to and from the cell state, enhancing the overall functionality of the network.

\begin{figure}[!ht]
\caption{Hidden units of an LSTM and a GRU network with specialized gates}
\includegraphics[width=\textwidth]{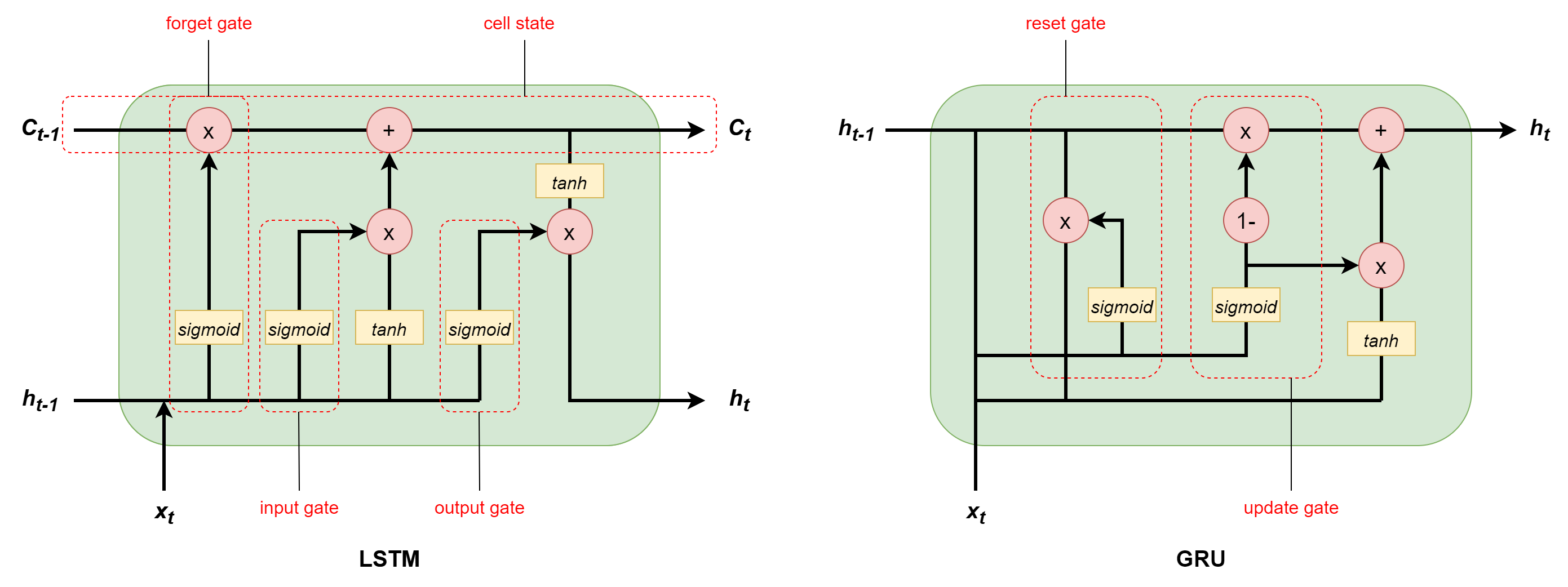}\\
\caption*{The figure displays the hidden units of LSTM and GRU networks compared to each other. The LSTM unit has three gates (forget, input, and output) and a cell state, distinguishing it from RNN. The cell state acts as a conveyor belt for retaining information over time, similar to human long-term memory, while the gates control information flow for improved network functionality. The GRU simplifies the LSTM structure by merging two gates into a single gate and combining the cell and hidden state. It replaces the output gate with an update gate that processes new information with respect to the previous hidden state and updates it accordingly.}
\end{figure}

Upon encountering information from the previous time step $h_{t-1}$ with new information $x_t$, the forget gate evaluates which information from the cell state $C_{t-1}$ should be discarded. Formally, this can be described as follows:

\begin{align*}
    f_t=\sigma(W_f\cdot[h_{t-1}\enspace x_t]+b_f). 
\end{align*}

This equation presents a concise representation of the mathematical operations within the forget gate. The terms enclosed in brackets represent the linear operations within the activation function, introducing nonlinearity. The input vector $[h_{t-1}\enspace x_t]$ is multiplied by the weight matrix $W_f$ and combined with the bias vector $b_f$, which is then passed through the activation function, typically a sigmoid function. This ensures that the output values range from 0 to 1. Later, these values will be multiplied by each corresponding element in the previous cell state $C_{t-1}$, determining the proportion of old information retained in the new cell state.

The subsequent step involves identifying the portion of new information considered valuable for storage in the cell state. This process occurs in two stages. Firstly, a node utilizing the hyperbolic tangent function proposes a vector of new candidate values for the cell state, denoted as $\tilde C_t$:

\begin{align*}
    \tilde C_t=\mathrm{tanh}(W_C\cdot[h_{t-1}\enspace x_t]+b_C). 
\end{align*}

Secondly, the input gate regulates the magnitude of the update, determining which values of $\tilde C_t$ will be stored in the cell state $C_t$:

\begin{align*}
    i_t=\sigma(W_i\cdot[h_{t-1}\enspace x_t]+b_i). 
\end{align*}

Similarly, the input gate generates values ranging from 0 to 1, which, when multiplied by $\tilde C_t$, determine the proportion of the new candidate that should be incorporated into the cell state:

\begin{align*}
    C_t=f_t\times C_{t-1} + i_t \times \tilde C_t.
\end{align*}

This equation provides a summary of the update process. The new cell state $C_t$ is obtained by combining the previous state $C_{t-1}$ multiplied by the forget gate $f_t$, which discards irrelevant information from the old state, and the new candidate values $\tilde C_{t}$ scaled by the input gate $i_t$, retaining only the most important information from the new candidate state.

The new hidden state $h_t$ is determined by the current cell state $C_t$ and the output gate $o_t$:

\begin{align*}
    o_t=\sigma(W_o\cdot[h_{t-1}\enspace x_t]+b_o). 
\end{align*}

The output gate in this equation utilizes a sigmoid activation function to determine the portion of the new cell state that will be outputted:

\begin{align*}
    h_t=o_t\times \mathrm{tanh}(C_t). 
\end{align*}

The cell state undergoes the hyperbolic tangent function and is multiplied by the output gate output, yielding the new hidden state $h_t$. Therefore, $h_t$ selectively captures the most pertinent information from the cell state, representing the short-term memory, while the cell state retains the long-term memory.

\subsubsection{Gated recurrent unit} 

In Figure 3, the right panel depicts a gated recurrent unit (GRU), which differs from the LSTM regarding the structure and quantity of its gates. GRU, introduced by \citet{cho2014}, aims to simplify the LSTM by combining two gates into a single gate and merging the cell state with the hidden state. Unlike LSTM, GRU eliminates the output gate and integrates the forget and input gates functions into a single gate known as the update gate. This update gate processes new information based on the previous hidden state, consequently updating the hidden state.

Initially, the previous hidden state $h_{t-1}$ and the current input $x_t$ are introduced and passed through the reset gate, which determines the extent to which past information should be disregarded:

\begin{align*}
    r_t=\sigma(W_r\cdot[h_{t-1}\enspace x_t]+b_r). 
\end{align*}

As the sigmoid activation function outputs values between 0 and 1, the resulting values can be interpreted as the proportion of past information to retain. This value, denoted as $r_t$, is multiplied by $h_{t-1}$ and passed through the hyperbolic tangent (tanh) activation function. This process yields a vector of new candidate values, represented as $\tilde h_t$:

\begin{align*}
    \tilde h_t=\mathrm{tanh}(W_h\cdot[r_t\cdot h_{t-1}\enspace x_t]+b_h). 
\end{align*}

The update gate governs the decision of which information to update in the next step:

\begin{align*}
    z_t=\sigma(W_z\cdot[h_{t-1}\enspace x_t]+b_z). 
\end{align*}

Finally, the new hidden state $h_t$ is then computed as a weighted average of the previous hidden state $h_{t-1}$ and the new candidate values $\tilde h_t$:

\begin{align*}
    h_t=(1-z_t)\cdot h_{t-1} + z_t\cdot \tilde h_t. 
\end{align*}

\subsection{Model specifications} 

Before estimating the model, it is necessary to prepare the chosen type of neural network. Figure 4 illustrates a simplified workflow chart outlining neural networks' forecasting process. Recession forecasting using neural networks involves five steps: Data must undergo preprocessing before being fed into the model. The network architecture is chosen depending on the specific type of neural network. Hyperparameters are optimized using the time series cross-validation in the expanding window scheme. The second and third steps can be combined, treating the number of layers and units within those layers as hyperparameters. Once the data is prepared and the optimal set of hyperparameters is determined, the model can be estimated, and predictions can be generated.

\begin{figure}[!ht]
\caption{Workflow chart of the forecasting process}
\includegraphics[width=\textwidth]{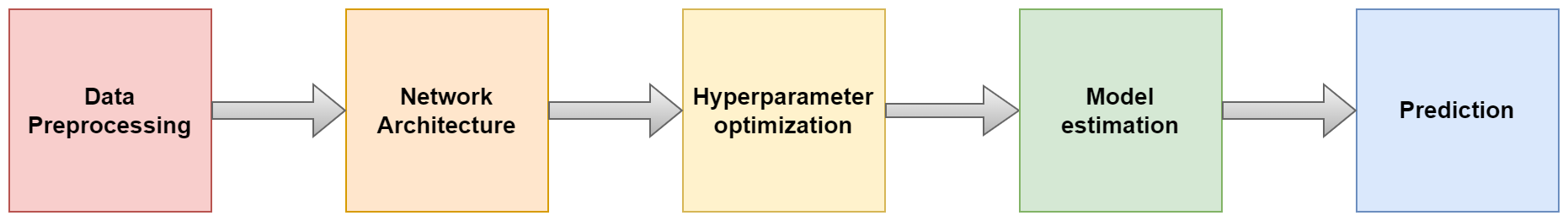}
\end{figure}

\paragraph*{Data preprocessing}

The first step involves preprocessing the data before it can be fed into the model. This preprocessing stage comprises multiple smaller steps. Initially, each series is evaluated and transformed individually to ensure stationarity. \citet{zhang2020} argue that while advanced neural network-based optimization algorithms can handle non-stationary data, they incorporate historical gradients in the update calculations, which can result in a lack of relevance to past information when dealing with non-stationary data where distributions change over time. Therefore, it is prudent to use stationary time series data. The second step involves seasonally adjusting and standardizing all explanatory variables. Lastly, a generator function generates data batches throughout the estimation process. Data batches are sets of data that are processed all at once. Optimization algorithms based on stochastic gradient descent often utilize smaller batch sizes or minibatches. \citet{bottou2010} and \citet{ge2015} highlight that minibatches offer the advantage of avoiding memory loss and introducing sufficient noise into each gradient update, aiding in escaping saddle points or local minima while achieving faster convergence. Additionally, the LSTM and GRU networks require a three-dimensional tensor as input, consisting of the batch size, the number of variables, and the number of time steps as axes. To accommodate the LSTM and GRU network, data is transformed into a suitable three-dimensional tensor format.

\paragraph*{Network architecture}

Section 3.1 presents three neural network models, each based on a different architecture. The number of variables determines the number of units in the input layer. The input dimension increases accordingly depending on the desired number of time points to consider. The number of hidden layers and units in those layers are treated as hyperparameters that should be optimized during cross-validation, which will be discussed in more detail later. Both the input and hidden layers of FFN use the rectified linear unit as the activation function for their units. The output layer consists of a single unit that uses the sigmoid activation function, producing a number between 0 and 1. This number can be interpreted as the probability of a recession.

As illustrated in Figure 2, the LSTM neural network architecture consists of as many LSTM cells as the number of time steps to look back. In the context of recession forecasting, I examine the temporal variations of the predictors over the last 12 months, corresponding to 12 lagged variables of the same predictor. \citet{deveaux1994} argue that due to the overparameterization of neural networks, individual weights associated with multicollinearity become less influential. The final output of a neural network results from various combinations of activation functions that involve interactions among predictors, making the impact of multicollinearity typically insignificant. Moreover, the backpropagation algorithm used in neural networks does not require inverting matrices, which can be problematic when perfect or severe multicollinearity exists. Hence, there are 12 LSTM cells, each with the same number of hidden units ranging from 16 to 64. The first unit in the first LSTM cell is linked to the first in the second LSTM cell, and the same holds for the rest of the LSTM cells. This approach ensures that each chain of units explores different variable dimensions at different time points. Additionally, the network may require a second chain of cells, leading to a hidden or stacked LSTM layer. The hyperparameter optimization process determines the optimal number of stacked layers. The output layer in LSTM is the same as in FFN. GRU follows the same architecture as LSTM but differs in the structure and functionality of the hidden units within the GRU network. 

\paragraph*{Hyperparameter optimization}

Deep neural networks with multiple hidden layers have the capacity to learn complex relationships. However, when training data are limited, some of these relationships may result from sampling noise and might not be present in real test data. This gives rise to overfitting, which occurs when a model fails to generalize from observed data to new data. Overfitting is a commonly acknowledged challenge in the supervised machine learning framework, and various methods employing different strategies have been devised to mitigate its effects and prevent the model from overfitting. \citet{ying2019} categorizes methods to address overfitting into three groups: network reduction, early stopping, and regularization. 

Network reduction involves reducing the depth and width of a neural network by decreasing the number of layers and units, thereby reducing the total number of parameters to estimate. This approach helps the model focus on capturing essential patterns in the training data, leading to a better generalization of test data. To implement network reduction, the network is limited to a maximum of one or two hidden layers, depending on the types of models, and a relatively small number of units is chosen for these layers. Additionally, through cross-validation, the model selects the optimal depth and width that minimizes validation loss, ensuring good generalization on unseen validation data. 

Early stopping is implemented to halt the training process when the gap between training loss and validation loss widens. The parameters are then set to the values corresponding to the smallest gap, preventing the model from memorizing the training data excessively. This helps in improving the model's performance on unseen data. To incorporate early stopping, a call is included during the estimation process that stops the iteration and restores the best parameters when the validation loss does not decrease for 5 consecutive epochs during cross-validation and for 10 consecutive epochs during the final estimation, where an epoch means one complete pass of the training dataset through the algorithm.

Regularization is a technique to mitigate the impact of less important variables. There are two commonly applied categories of regularization methods in neural networks. The first category involves adding a penalty term or regularizer to the loss function. Two well-known types of regularizers, namely $L1$ and $L2$, are commonly used, similar to their application in penalized regression. $L1$ regularization, known as Lasso regression (\citet{tibshirani1996}), assigns zero weights to unimportant variables, effectively removing them from the model. This ensures that only influential variables with significant effects on the variable of interest are retained. On the other hand, $L2$ regularization, known as Ridge regression (\citet{hoerl1970}), assigns lower weights to unimportant variables rather than discarding them completely. This approach aims to extract as much relevant information as possible while controlling model complexity and reducing overfitting. The second category of regularization methods involves using a technique called dropout. Dropout was introduced by \citet{srivastava2014} and involves randomly dropping units and their connections from the neural network during training. A certain percentage of hidden units are randomly dropped to create a thinned network, which is then trained using stochastic gradient descent. After training, the dropped units are restored, and the process is repeated. Dropout approximates the effects of averaging the estimates over multiple smaller networks while preventing overfitting. I implement the dropout technique in my model, treating the dropout percentage as a hyperparameter to be optimized through cross-validation.

Besides the hyperparameters related to the overfitting issue, neural networks also have other hyperparameters that must be determined prior to training. These hyperparameters include the number of layers, the number of units in these layers, batch size, and learning rate. The values of these hyperparameters must be specified before training can begin. There are various methods to search for the best combination of hyperparameters, known as hyperparameter optimization. A straightforward approach is a manual search, where the researchers select their own set of hyperparameters based on existing literature and evaluate the model performance on validation data. This process is repeated with different hyperparameter settings until the best combination is found. However, manual search can be time-consuming and does not guarantee finding the optimal hyperparameters. Grid search is another approach where a predetermined set of values is specified for each hyperparameter. Every possible combination of values is then evaluated, resulting in many trials. However, as indicated by \citet{bellman1961}, the number of possible combinations grows exponentially with the number of hyperparameters, leading to the curse of dimensionality. This can make grid search impractical for models with many hyperparameters. A more effective approach, demonstrated by \citet{bergstra2012}, is random search. The random search involves randomly selecting values from predetermined ranges for each hyperparameter. The advantage of random search is that it can identify a good or even better set of hyperparameters within a smaller fraction of computation time compared to manual or grid search methods. Following this approach, the random search is conducted using predefined ranges of values for the hyperparameters. The types of hyperparameters and their specific ranges used for random search in the optimization process are reported in Table 3.

\begin{table}[!ht]
    \caption{Tuning hyperparameters} 
    \begin{tabular}{l|l}
    \hline
    Type & Range \\
    \hline
    \# hidden layer & $\in\{1,2\}$ \\
    \# unit & $\in\{16,32,64\}$ \\
    batch size & $\in\{16,32,64\}$ \\
    dropout & $\in\{0,0.1,0.2,0.3,0.4,0.5\}$ \\
    recurrent dropout & $\in\{0,0.1,0.2,0.3,0.4,0.5\}$ \\
    weight decay & $\in\{0,0.1,0.2\}$ \\
    learning rate & $\in\{0.01,0.001\}$ \\
    \hline
    \end{tabular} \\
    \caption*{The table provides the ranges of possible values for different neural network-based hyperparameters of the neural networks used in the analysis. The optimal combination of these hyperparameters is obtained by random search using the time series cross-validation technique in an expanding window scheme.}
\end{table}

Figure 5 illustrates the form of cross-validation employed to determine the optimal set of hyperparameters. Each block of the training set is divided into two sections at every iteration, with the validation split always coming after the training split. This approach ensures that the natural order of observations is maintained. For each block, the size of the validation set remains the same, while the training set gets larger as the validation set of the prior block is added to it. The number of iterations in the cross-validation setup depends on the size of each block. If the block size is small, the number of blocks and iterations increases. However, selecting an excessively small block size may result in certain blocks having insufficient or no recession data, leading to sampling bias. Considering the information presented in Table 2, which indicates the longest period between two recessions in US history prior to the Great Recession as 128 months, the length of each validation block is set to be 128 months to ensure that each validation set contains data from at least one recession. The length $l$ of the first training block is equal to $l=N_{total}-128\times(\floor*{\frac{n_{total}}{128}}-1)$, where $N_{total}$ denotes the total number of observations in a data vintage. This guarantees that the first training block is always larger than the validation block. Subsequently, an expanding window approach is adopted, increasing the length of the training set by 128 months at each step while maintaining a validation block of the same length (128 months). This methodology ensures that scarce recession data are fully utilized.

\begin{figure}[!ht]
\caption{Cross-validation for time series}
\includegraphics[width=\textwidth]{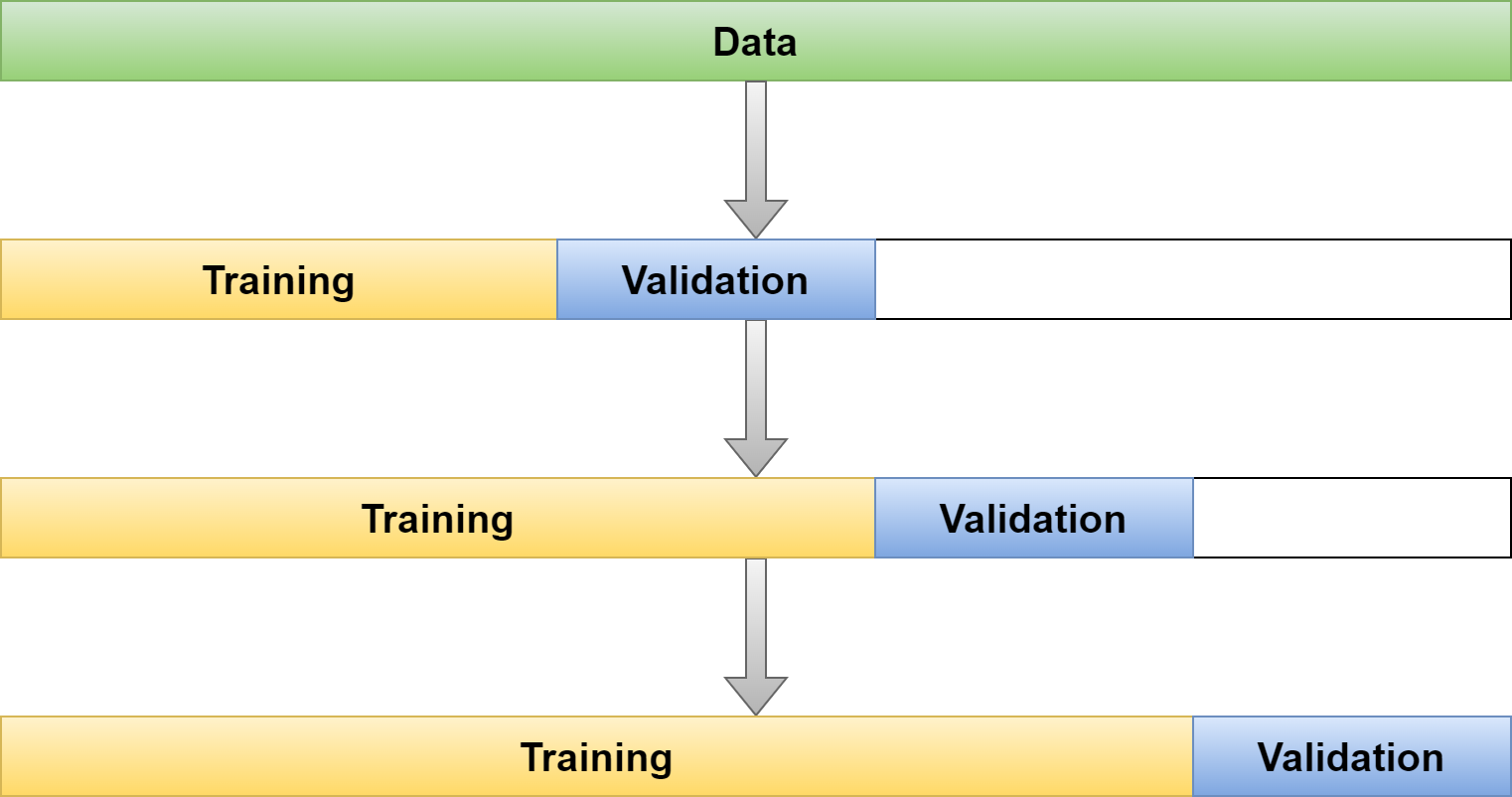}
\caption*{The figure illustrates the form of cross-validation. Each block of the training set is divided into two parts, and the training set is always before the validation set to preserve the natural order of observations within each block. The size of the validation set remains constant for each block, while the training set grows larger as the prior block's validation set is incorporated into it.}
\end{figure}

The average validation loss across all cross-validations is computed and stored for each combination of hyperparameters. The optimal choice is the set of hyperparameters with the lowest average validation loss. According to mathematical calculations by \citet{bergstra2012}, using a random search with 60 trials will likely yield a set of hyperparameters within the top 5\% interval around the optimal solution with a 95\% probability. Therefore, 60 trials are conducted for each training process. Once the optimal combination of hyperparameters is determined, the entire dataset is split into training and validation parts in the same proportion, maintaining the chronological order of the data. Afterward, the final estimation and validation of the model are performed to improve the accuracy of the parameters and check for any signs of overfitting.  

\paragraph*{Model estimation}

Once the hyperparameters are optimized, the neural network is prepared to fit the training data. The data is fed into the model, processed, and transformed into an output. This output is then compared to the actual output, and the error signals are propagated backward through the network to adjust the parameters accordingly. This iterative process is known as backpropagation and continues until a specific condition is satisfied. Although the concept appears straightforward, the estimation process involves extensive mathematical computations. The parameter values are estimated to minimize the following binary cross-entropy loss function:

\begin{align*}
    \mathrm{Loss}=\frac{1}{n}\sum_{i=1}^N -[y_i\cdot \mathrm{ln}(p_i)+(1-y_i)\cdot \mathrm{ln}(1-p_i)]. 
\end{align*}

In most cases, no analytical solution is available for minimizing such problems, so the parameters must be estimated numerically. I utilize the Adam optimizer introduced by \citet{kingma2014} to perform this estimation. The Adam optimizer is designed to optimize stochastic loss functions using first-order gradients. It combines the strengths of two popular optimization methods, AdaGrad (\citet{duchi2011}) and RMSProp (\citet{tieleman2012}). The Adam optimizer offers computational efficiency and requires minimal memory resources.

\paragraph*{Prediction}

Since the effectiveness of predictive variables may vary across different forecast horizons, I explore five distinct windows: nowcasting, immediate-term, short-term, medium-term, and long-term. These windows correspond to predicting the current month, one month ahead, three months ahead, six months ahead, and twelve months ahead, respectively.

After the models generate probability predictions, they are transformed into binary indicators that represent the state of the economy, recession, or boom based on predefined cutpoints. The conventional approach uses a fixed threshold, typically set at 0.5. Under this approach, if a probability prediction equals or exceeds 0.5, it is classified as a recession, and vice versa. However, it is often mentioned that business cycles exhibit asymmetry, and therefore, a 50\% threshold may not be optimal for linear modeling methods. \citet{berge2011} propose an optimal cutpoint within the range of 0.3 to 0.6 based on smoothed state probabilities estimated by \citet{chauvet2008}. \citet{ng2014} suggests using different thresholds for different forecast horizons, ranging from 0.3 to 0.44. \citet{vrontos2021} adopt a fixed threshold of 0.33 for classification and evaluation purposes. In the case of neural networks, they are assumed to be capable of capturing the asymmetry and other nonlinear characteristics of business cycles. Furthermore, they must be compared to linear frameworks to assess their predictability. Hence, for comparison purposes, I follow the traditional approach and use a fixed threshold of 0.5 for all the models in this study.

The model is reestimated for the entire out-of-sample period to capture dynamic structural changes in the data. To manage computational resources, a forecast window of 12 months is selected, during which the estimated parameters remain constant. This means that once a neural network model is built, it is used to predict for the next 12 months using the latest available data. Afterward, the observed data for those 12 months is added to the existing sample, and the neural network model is reestimated using this extended dataset. The updated parameters are then used to forecast for the following 12 months. Although this repeated estimation approach is time-consuming and costly, it mimics the real-time process of generating predictions using the most up-to-date data available. This expanding window procedure is repeated, gradually increasing the size of the training and validation set to generate out-of-sample forecasts for the test period spanning from November 2006 to October 2021. The same approach is adopted for the other forecast horizons. 

\subsection{Performance evaluation}

The estimated models produce two types of forecasts: probability and point predictions. The models first generate probability predictions. The next step transforms these probabilities into binary point predictions using a specified threshold. While the second type may not always be necessary, it is included to ensure the forecast is more easily interpretable for the end users. The performance of both the probability and point predictions is evaluated using various statistical measures. Except for metrics related explicitly to probability predictions, the evaluation is based on the output of a contingency table, commonly known as a confusion matrix, as illustrated in Table 4. This matrix provides a framework for assessing the accuracy of the point predictions.

\begin{table}[!ht]
        \caption{Confusion Matrix}
    \begin{tabular}{c|c|c|c}
    \hline
        \multicolumn{2}{c|}{\multirow{2}{*}{}} & \multicolumn{2}{|c}{Predicted} \\
        \cline{3-4} 
        \multicolumn{2}{c|}{} & Positive & Negative \\
        \hline
        \multirow{2}{*}{Actual} & Positive & TP & FN \\
         & Negative & FP & TN \\
         \hline
    \end{tabular}\\
    \caption*{The table contains elements representing the count of observations belonging to each category. True positives (TP) and true negatives (TN) indicate the correct classification of positive and negative outcomes. False positives (FP) occur when the prediction is positive, but the actual value is negative. Conversely, false negatives (FN) arise when the prediction is negative, but the actual value is positive.} 
\end{table} 

The confusion matrix comprises elements representing the count of observations belonging to different categories. True positives (TP) and true negatives (TN) indicate the correct classification of positive and negative outcomes. False positives (FP) occur when the prediction is positive while the actual value is negative. Conversely, false negatives (FN) arise when the prediction is negative while the actual value is positive. By utilizing these counts, various performance evaluation measures are derived to assess the overall performance of the models.

\begin{table}[!ht]
    \caption{Performance evaluation metrics} 
    \resizebox{\textwidth}{!}{\begin{tabular}{c|l|c}
    \hline
    Type & Metric & Formula \\
    \hline
    Probability & Area under the ROC curve & $\int_0^1 ROC(c)\,dc$ \\  
    Prediction & Area under the PR curve & $\int_0^1 PR(c)\,dc$\\ 
    \hhline{=|=|=}
     & Sensitivity & $\frac{TP}{TP+FN}$\\  
     & Specificity & $\frac{TN}{FP+TN}$\\ 
     Point & Precision & $\frac{TP}{TP+FP}$ \\ 
     Prediction & Balanced accuracy & $\frac{Sensitivity+Specificity}{2}$\\  
     & Matthews correlation coefficient & \begin{tabular}{c}$\frac{TP\times TN-FP\times FN}{\sqrt{(TP+FP)(TP+FN)(TN+FP)(TN+FN)}}$\end{tabular}\\  
     & $F_1$-Score & $\frac{2}{\frac{1}{Sensitivity}+\frac{1}{Precision}}$\\  
    \hline
    \end{tabular}}
     \caption*{The table reports the metrics used for the performance evaluation of a forecasting model. They are divided into two groups, depending on which type of predictions they refer to.} 
\end{table}

Table 5 reports the list of the metrics for the performance evaluation. A carefully selected set of statistical metrics is used to assess the performance from different perspectives. The metrics, Area Under the Receiver Operating Characteristic Curve (AUROC) and Area Under the Precision-Recall Curve (AUPRC), are employed to measure the models' raw performance in the sense that these metrics do not require any threshold to convert probabilities into binary outcomes, making them suitable for evaluating the models' pure performance. The Receiver Operating Characteristics (ROC) curve displays the entire set of possible combinations of true positive rates $TPR(c)=\frac{TP(c)}{\#\:Actual\:Positive}$ and false positive rates $FPR(c)=\frac{FP(c)}{\#\:Actual\:Negative}$ for some cutpoint $c\in(0,1)$ that maps the predicted probability to a binary category. Similarly, the Precision Recall (PR) curve plots the complete set of possible combinations of precision and recall for $c\in(0,1)$, where recall is a synonym for sensitivity. The Area under the ROC and PR curves increases with the underlying metrics for a given cutpoint. By aggregating over the entire set of cutpoints, these curves deliver a framework to assess the pure predictive ability of a forecasting model. \citet{tharwat2021} provides a comprehensive overview of other metrics and discusses their strengths and weaknesses, particularly in imbalanced binary classification problems. 

\section{Empirical results}

\renewcommand{\arraystretch}{0.74}
\begin{table}[!ht]
    \caption{Performance evaluation measures: A real-time assessment} 
    \begin{adjustbox}{width=\textwidth}
    \begin{tabular}{lrrrrrrrr}
    \hline
    Method & AUROC & AUPRC & BAcc & MCC & $F_1$-Score & Sensitivity & Specificity & Precision \\
    \hline
    \multicolumn{9}{l}{Panel A: nowcasting setup} \\
    \hline
    Logit & 0.853 & 0.365 & 0.812 & 0.496 & 0.546 & 0.750 & 0.874 & 0.429 \\
    Ridge & 0.920 & 0.529 & 0.875 & 0.597 & 0.630 & 0.850 & 0.893 & 0.500 \\
    FFN & 0.917 & 0.642 & 0.906 & 0.668 & 0.692 & 0.900 & 0.912 & 0.563 \\
    LSTM & 0.899 & 0.754 & 0.931 & 0.797 & 0.818 & 0.900 & 0.962 & 0.750\\
    GRU & 0.890 & 0.837 & 0.928 & 0.778 & 0.800 & 0.900 & 0.956 & 0.720 \\
    \hline
    \multicolumn{9}{l}{Panel B: immediate-term setup} \\
    \hline
    Logit & 0.671 &  0.243 & 0.634 & 0.230 & 0.327 & 0.400 & 0.868 & 0.276\\
    Ridge & 0.931 & 0.555 & 0.881 & 0.634 & 0.667 & 0.850 & 0.912 & 0.548 \\
    FFN & 0.860 & 0.652 & 0.825 & 0.623 & 0.667 & 0.700 & 0.950 & 0.636 \\
    LSTM & 0.847 & 0.514 & 0.890 & 0.607 & 0.632 & 0.900 & 0.881 & 0.487\\
    GRU & 0.896 & 0.831 & 0.944 & 0.887 & 0.900 & 0.900 & 0.987 & 0.900 \\
    \hline
    \multicolumn{9}{l}{Panel C: short-term setup} \\
    \hline
    Logit & 0.524 & 0.170 & 0.587 & 0.156 & 0.261 & 0.300 & 0.874 & 0.231\\
    Ridge & 0.928 & 0.522 & 0.850 & 0.575 & 0.615 & 0.800 & 0.899 & 0.500 \\
    FFN & 0.872 & 0.800 & 0.841 & 0.732 & 0.757 & 0.700 & 0.981 & 0.824 \\
    LSTM & 0.858 & 0.395 & 0.784 & 0.508 & 0.565 & 0.650 & 0.918 & 0.500\\
    GRU & 0.885 & 0.544 & 0.931 & 0.797 & 0.818 & 0.900 & 0.962 & 0.750 \\
    \hline
    \multicolumn{9}{l}{Panel D: medium-term setup} \\
    \hline
    Logit & 0.632 & 0.174 & 0.515 & 0.029 & 0.143 & 0.150 & 0.881 & 0.136 \\
    Ridge & 0.910 & 0.423 & 0.840 & 0.541 & 0.582 & 0.800 & 0.881 & 0.457 \\
    FFN & 0.795 & 0.496 & 0.731 & 0.510 & 0.556 & 0.500 & 0.962 & 0.625 \\
    LSTM & 0.853 & 0.519 & 0.903 & 0.655 & 0.679 & 0.900 & 0.906 & 0.546\\
    GRU & 0.731 & 0.250 & 0.647 & 0.266 & 0.356 & 0.400 & 0.893 & 0.320 \\
    \hline
     \multicolumn{9}{l}{Panel E: long-term setup} \\
    \hline
    Logit & 0.737 & 0.250 & 0.625 & 0.235 & 0.326 & 0.350 & 0.899 & 0.304 \\
    Ridge & 0.818 & 0.388 & 0.668 & 0.265 & 0.357 & 0.500 & 0.837 & 0.278 \\
    FFN & 0.856 & 0.365 & 0.612 & 0.235 & 0.316 & 0.300 & 0.925 & 0.333 \\
    LSTM & 0.862 & 0.371 & 0.725 & 0.396 & 0.468 & 0.550 & 0.899 & 0.407\\
    GRU & 0.909 & 0.674 & 0.837 & 0.707 & 0.737 & 0.700 & 0.975 & 0.778 \\
    \hline
    \end{tabular}
    \end{adjustbox}
     \caption*{The table reports the performance evaluation measures of forecasts obtained by logit models, ridge logit models, feed-forward (FFN), long short-term memory (LSTM), and gated recurrent unit (GRU) neural networks over different forecast horizons for the out-of-sample period, November 2006 to October 2021: Panel (A) presents the nowcasts. Panel (B), (C), (D), and (E) display the 1-month-ahead forecasts, the 3-month-ahead forecasts, the 6-month-ahead forecasts, and the 12-month-ahead forecasts, respectively.}
\end{table}

Table 6 presents the out-of-sample forecast performance of the models for various forecast horizons. In the nowcasting setup, there is a difference in predictive performance between the types of logit models and neural network models. The standard logit model does not outperform other models in any considered metric, although the difference is minor compared to other forecast horizons. The ridge logit model performs similarly to FFN. Although FFN performs slightly better, the differences are small for most metrics. However, LSTM and GRU perform significantly better in MCC and $F_1$-Score, mainly due to their high precision values. For instance, LSTM correctly identifies 90\% of recession months, while 75\% of its positive predictions are correct. In contrast, the ridge logit model has a precision of 51.5\%, indicating that LSTM accurately predicts recessions without generating too many false alarms. The AUPRC values also highlight the superiority of LSTM and GRU models over the others, with GRU achieving an AUPRC of 0.837 compared to 0.642 for FFN and 0.529 for the ridge logit model. \citet{davis2006} argue that PR curves give a more informative picture of an algorithm's performance when dealing with imbalanced datasets. The ratios of the number of months in booms to the number of months in recessions in 194 real-time vintages range from 4.5 to 6.7 with a mean value of 6, which means that the period of booms is, on average, six times longer than that of recessions. The percentage of recessions in the datasets fluctuates around 0.2 across the vintages. Although not highly skewed, the datasets are certainly imbalanced, and the difference in AUPRC values suggests that LSTM and GRU are better at extracting hidden patterns from scarce recession data.  

Moving to the immediate-term setup, logit models perform significantly worse, with MCC and $F_1$-Score approximately 40\% lower than before. Conversely, the other models show similar or slightly improved performance. GRU exhibits the highest performance across various summarizing metrics such as balanced accuracy, MCC, and $F_1$-Score. Although the ridge logit model demonstrates better predictive performance in the short-term setup than LSTM, neural network models, particularly GRU, outperform the other models on average. GRU maintains a 90\% accuracy in classifying recession months with higher specificity and precision, resulting in higher values of balanced accuracy, MCC, and $F_1$-Score. In the medium-term setup, LSTM performs slightly better than the other models, while GRU performs poorly, and thus, the difference between the two groups is minimal. However, LSTM achieves 90\% sensitivity and over 90\% specificity while maintaining precision above 0.5. The overall average performance of the models is the lowest among different forecast horizons. In the long-term setup, GRU surpasses the other models by a significant margin, with an $F_1$-Score of 0.737 compared to 0.468 for LSTM and 0.357 for the ridge logit model. GRU still accurately predicts 70\% of recession months, with 77.8\% of its positive predictions being correct.

\begin{figure}[!ht]
    \caption{Receiver Operating Characteristic (ROC) curves}
    \centering
    \resizebox{.83\linewidth}{!}{
    \begin{tabular}{c c}
        (A) &  (B) \\ 
        \includegraphics[width=.43\textwidth]{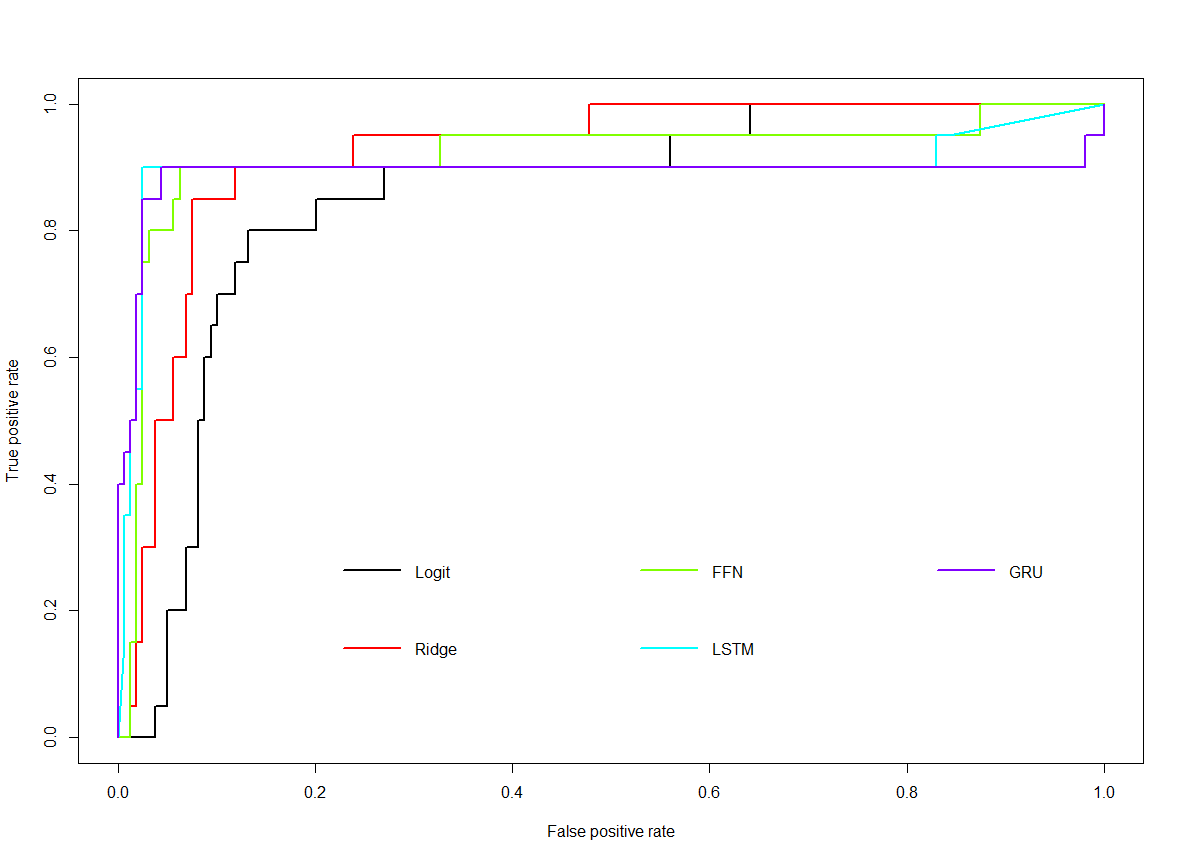} &  \includegraphics[width=.43\textwidth]{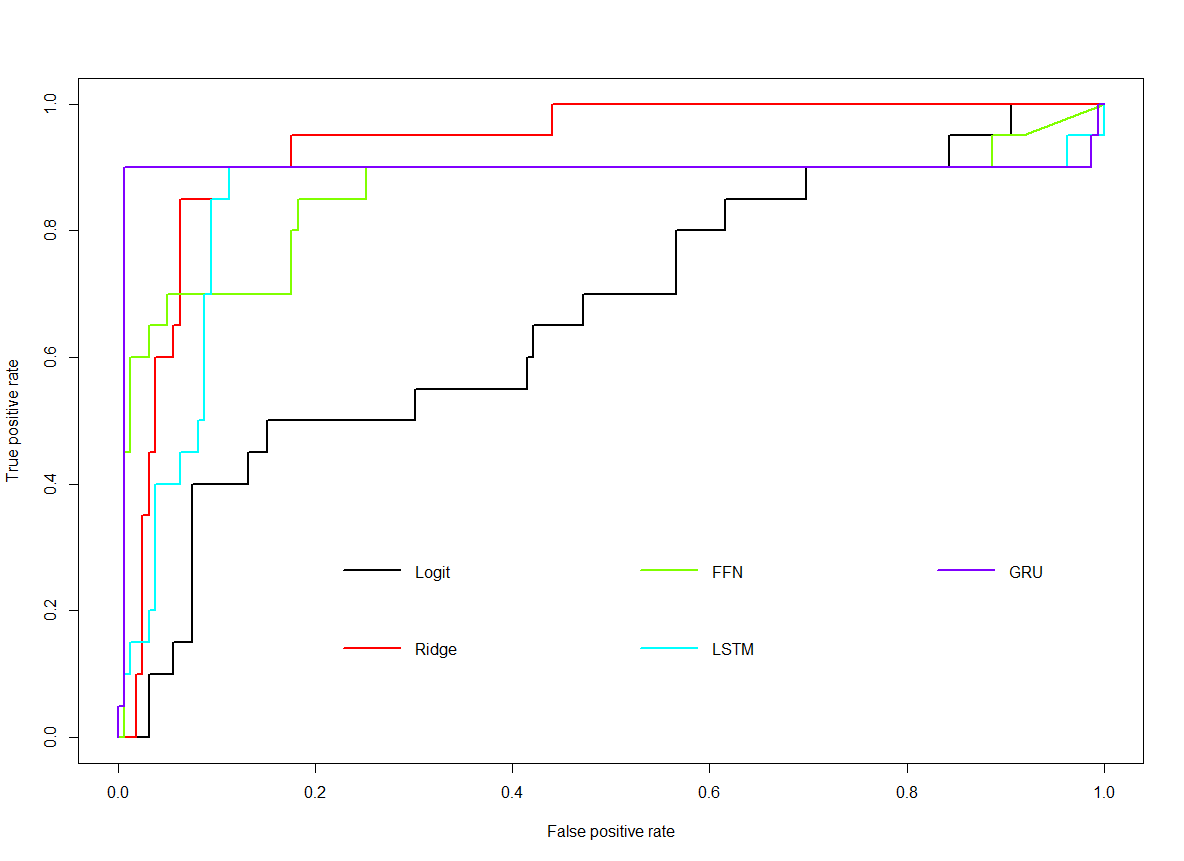} \\
        (C) & (D) \\
        \includegraphics[width=.43\textwidth]{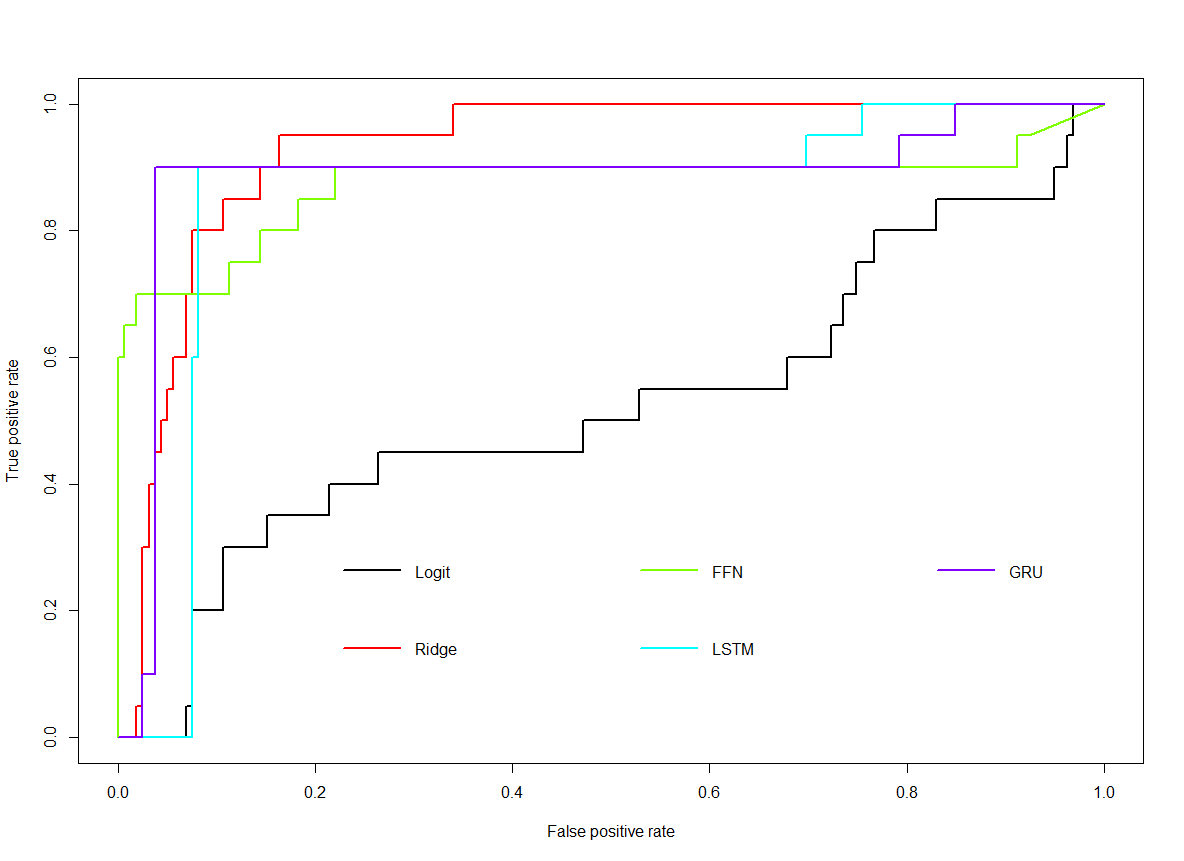} & \includegraphics[width=.43\textwidth]{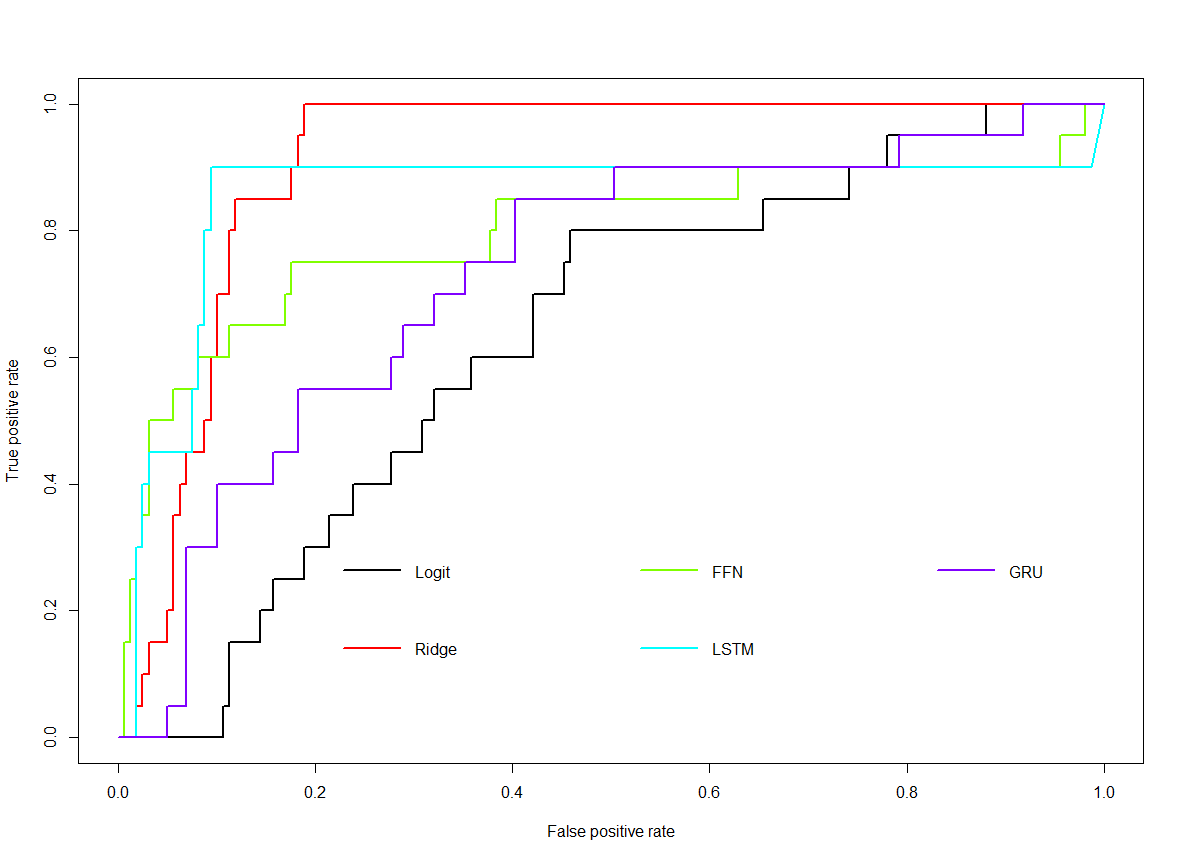} \\
        (E) & \\
        \includegraphics[width=.43\textwidth]{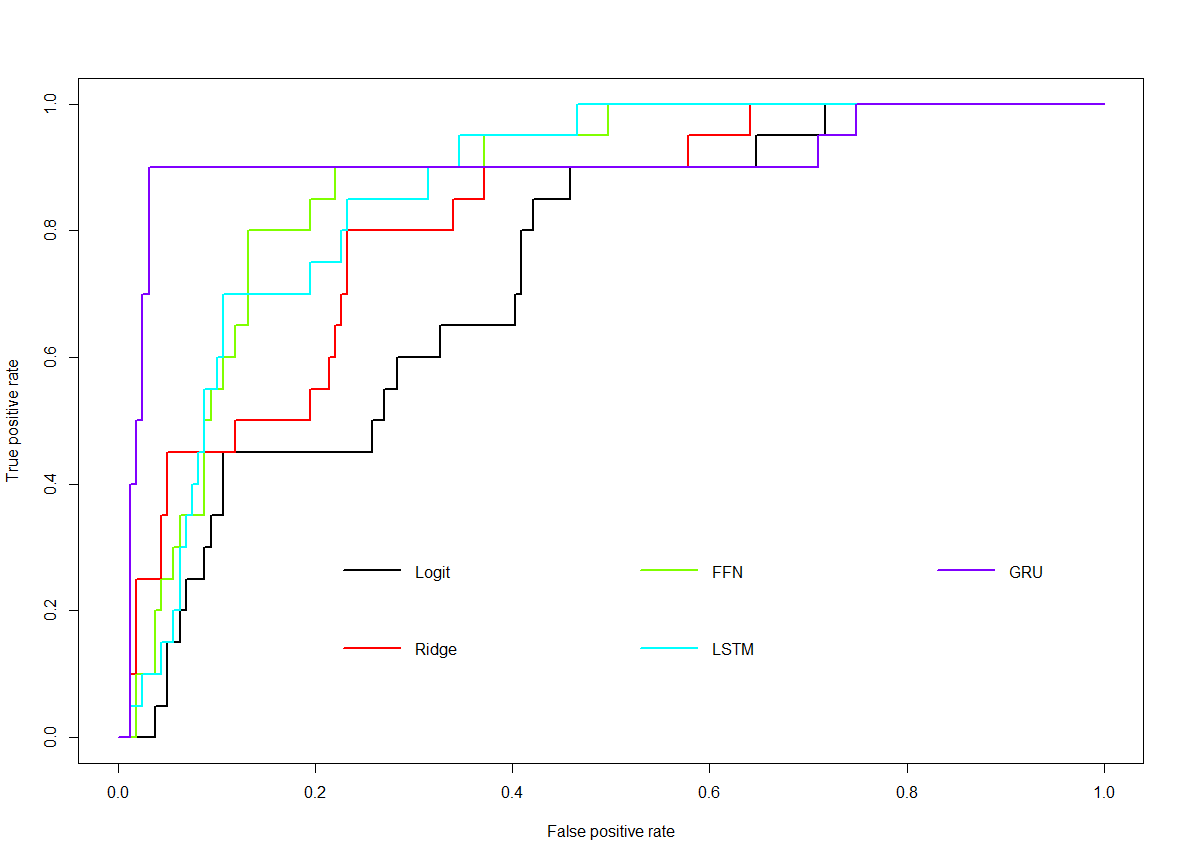} & \\
    \end{tabular}}
    \caption*{The figure illustrates receiver operating characteristic (ROC) curves for the out-of-sample period from November 2006 to October 2021: Panel (A) shows the ROC curves for the nowcasting forecast horizon, comparing the logit model, ridge logit model, FFN, LSTM, and GRU. Panels (B), (C), (D), and (E) present the ROC curves for the immediate-term (1-months-ahead), short-term (3-months-ahead), medium-term (6-months-ahead), and long-term (12-months-ahead) forecast horizon, respectively.}
\end{figure}

\begin{figure}[!ht]
    \caption{Precision Recall (PR) curves}
    \centering
    \resizebox{.83\linewidth}{!}{
    \begin{tabular}{c c}
        (A) &  (B) \\
        \includegraphics[width=.43\textwidth]{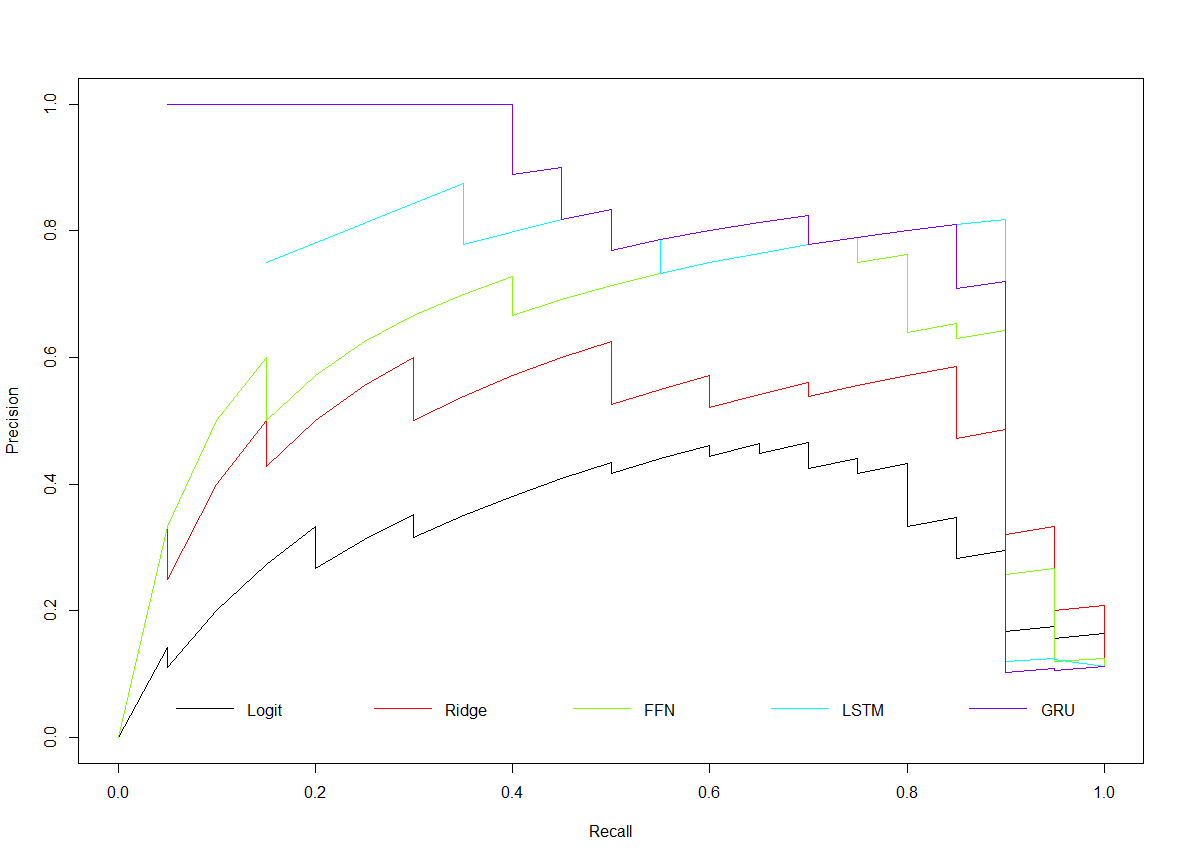} &  \includegraphics[width=.43\textwidth]{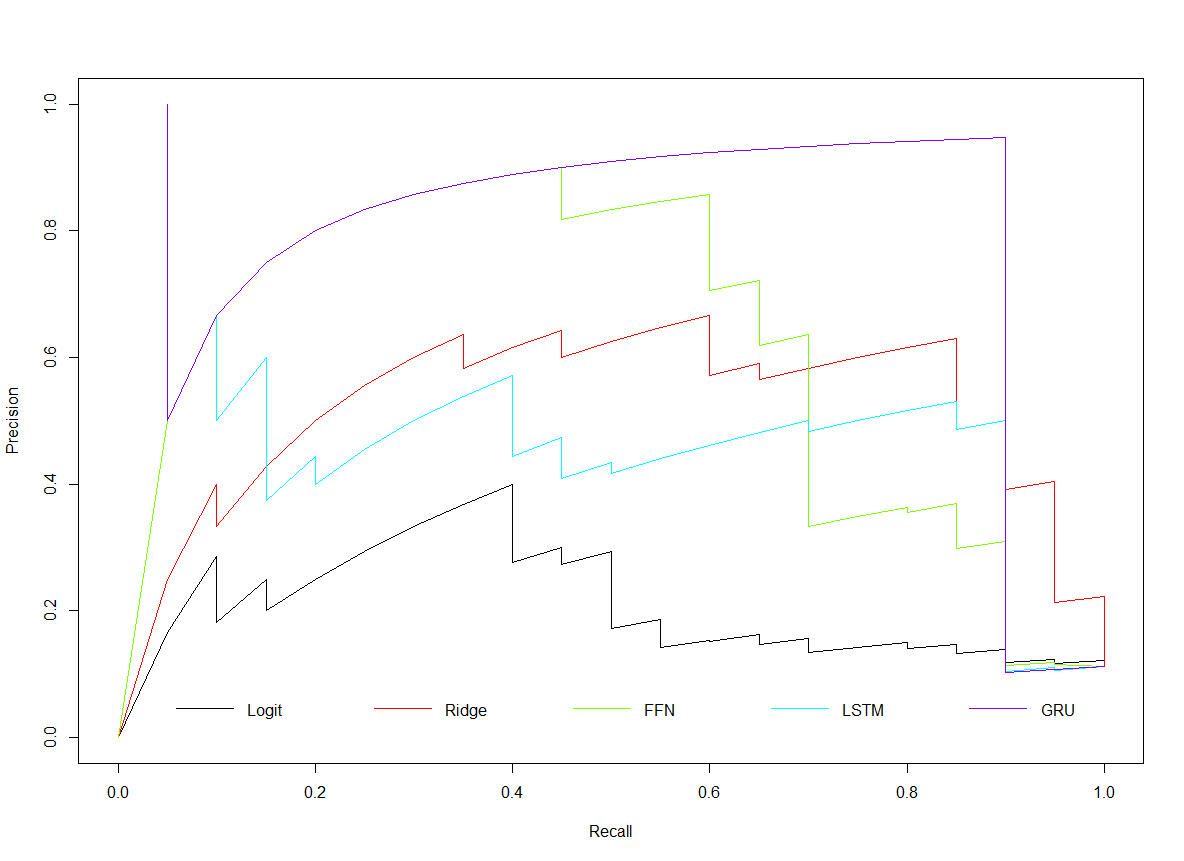} \\
        (C) & (D) \\
        \includegraphics[width=.43\textwidth]{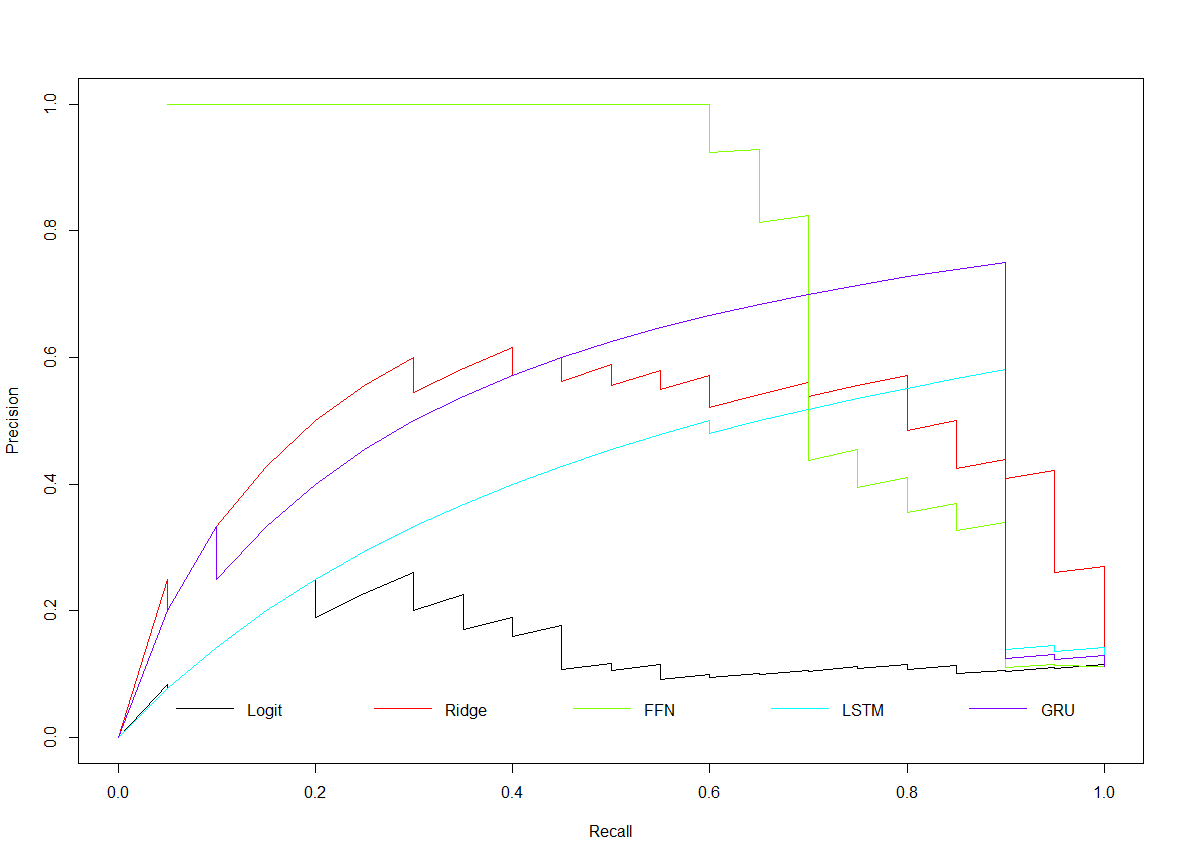} & \includegraphics[width=.43\textwidth]{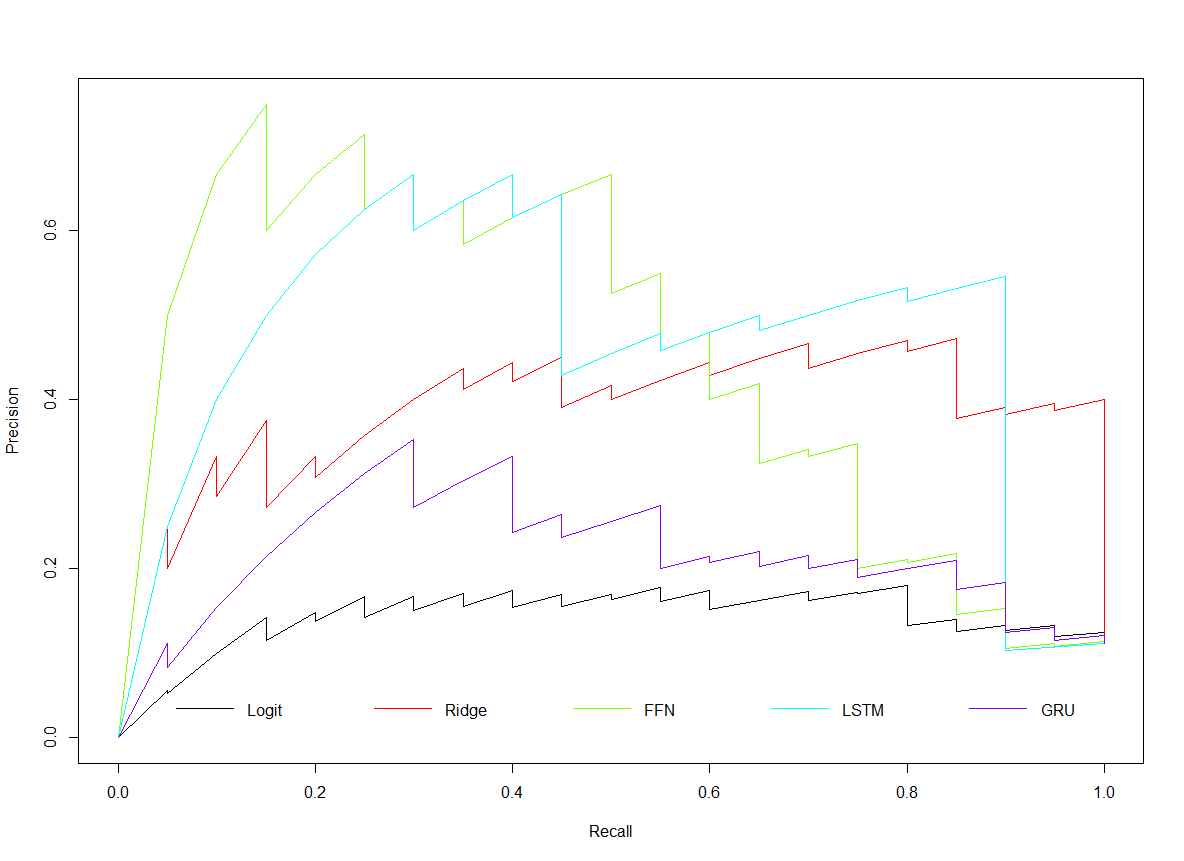} \\
        (E) & \\
        \includegraphics[width=.43\textwidth]{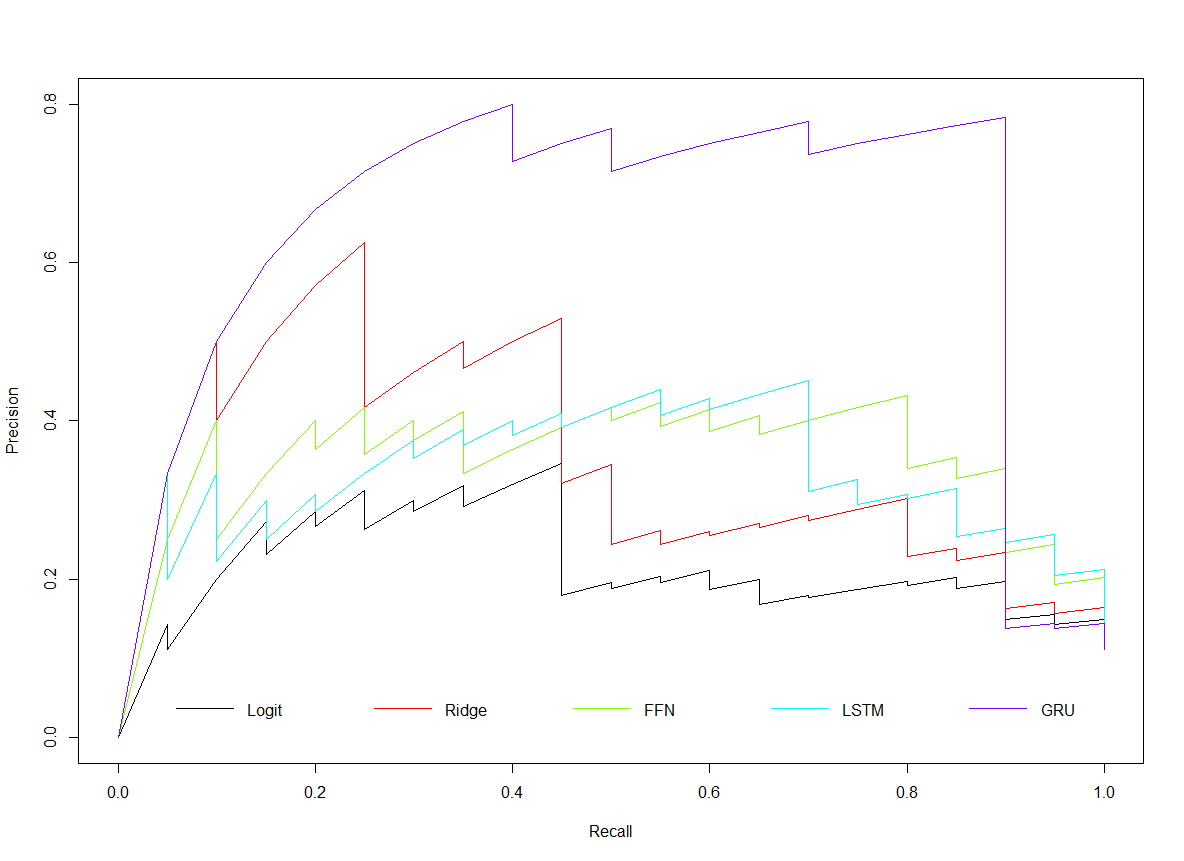} & \\
    \end{tabular}}
    \caption*{The figure depicts precision-recall (PR) curves for the out-of-sample period from November 2006 to October 2021: Panel (A) shows the PR curves for the nowcasting forecast horizon, comparing the logit model, ridge logit model, FFN, LSTM, and GRU. Panels (B), (C), (D), and (E) present the PR curves for the immediate-term (1-months-ahead), short-term (3-months-ahead), medium-term (6-months-ahead), and long-term (12-months-ahead) forecast horizon, respectively.}
\end{figure}

Neural network models, especially LSTM and GRU, significantly improve predictive performance compared to logit models. Across all metrics considered, neural network models consistently outperform logit models by a significant margin, except for the nowcasting setup. This superiority is evident in pure performance metrics, such as AUROC and AUPRC, and threshold-based point prediction metrics. The disparity is particularly prominent in AUPRC, a more reliable indicator when dealing with imbalanced data, highlighting the neural network models' ability to handle class imbalance effectively. The performance difference decreases when comparing them to ridge logit models. Ridge logit models exhibit higher AUROC values than other model specifications, except for the long-term setup. However, regarding AUPRC, neural networks again prove to be a better choice. On average, neural network models perform better than ridge logit models. The gap widens further when considering only the two recurrent neural network models. Depending on the forecast horizons, LSTM or GRU may lead the competition, with occasionally substantial differences between them and the other models. 

The ROC and PR curves of the models for five different forecast horizons are presented in Figures 6 and 7, respectively. Panel B and C of the ROC curves highlight the noticeable difference in forecast performance between the logit model (represented by the black line) and the other models. In Panel D and E, although the lines are closer together, there is still some discernible gap between each model. Overall, the ridge logit model, LSTM, and GRU exhibit superior performance of probability predictions across various forecast horizons. Figure 7 demonstrates that the green, blue, and purple lines lie, on average, above the red line and, even more so, above the black line. This finding supports the argument that, compared to linear models, neural network models are more capable of handling class imbalance in the data.

\begin{figure}[!ht]
    \caption{Out-of-sample recession probabilities}
    \centering
    \resizebox{.76\linewidth}{!}{
    \begin{tabular}{c c}
        (A) &  (B) \\
        \includegraphics[width=.39\textwidth]{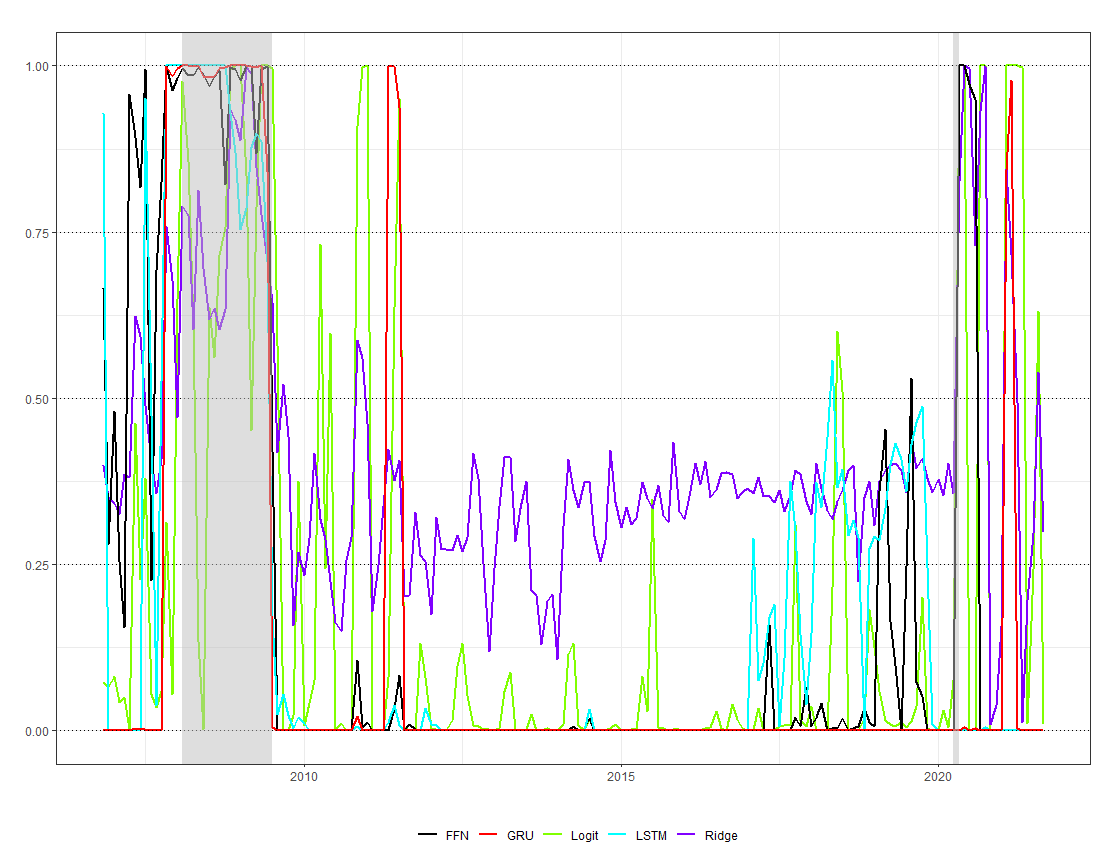} &  \includegraphics[width=.39\textwidth]{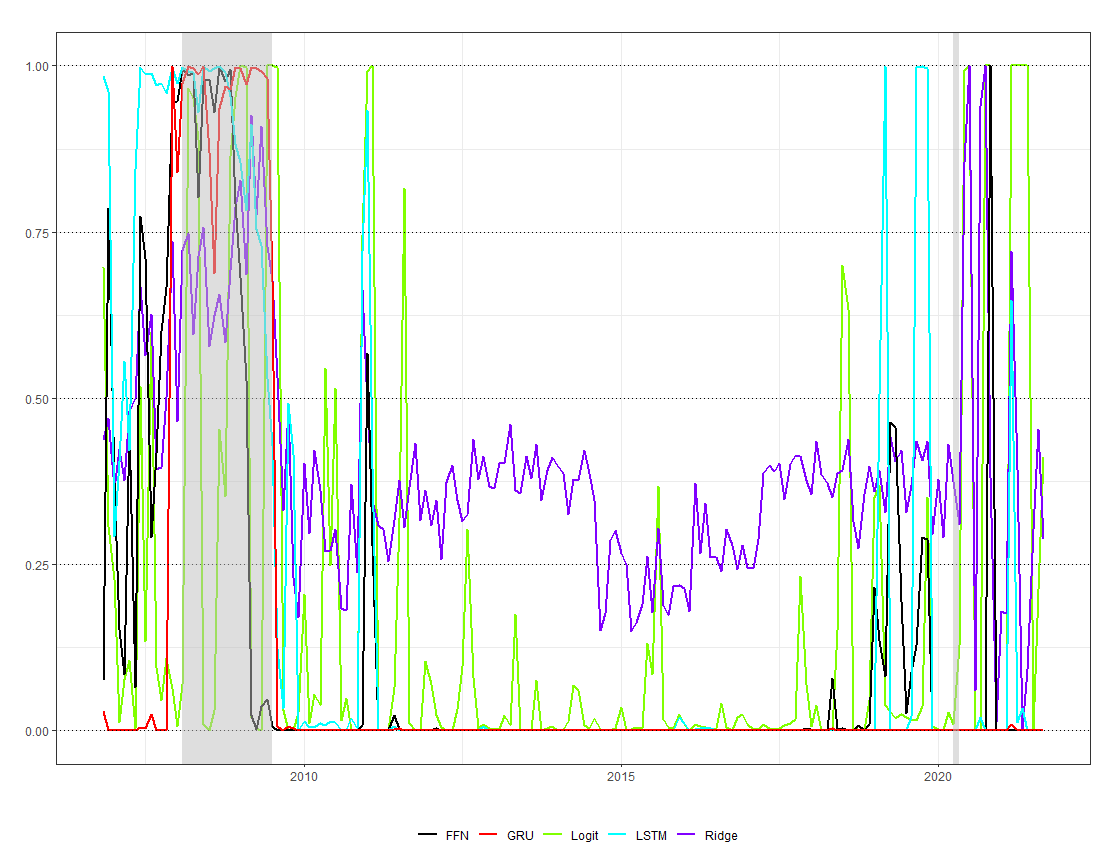} \\
        (C) & (D) \\
        \includegraphics[width=.39\textwidth]{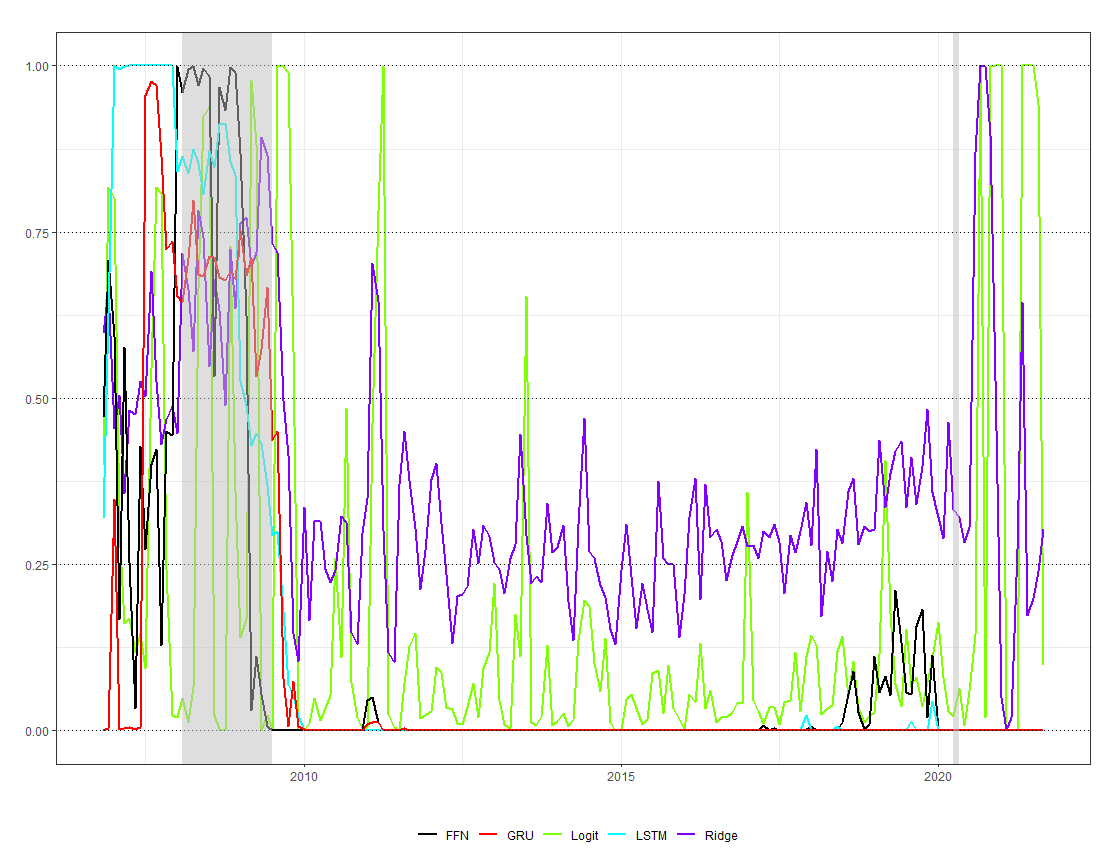} & \includegraphics[width=.39\textwidth]{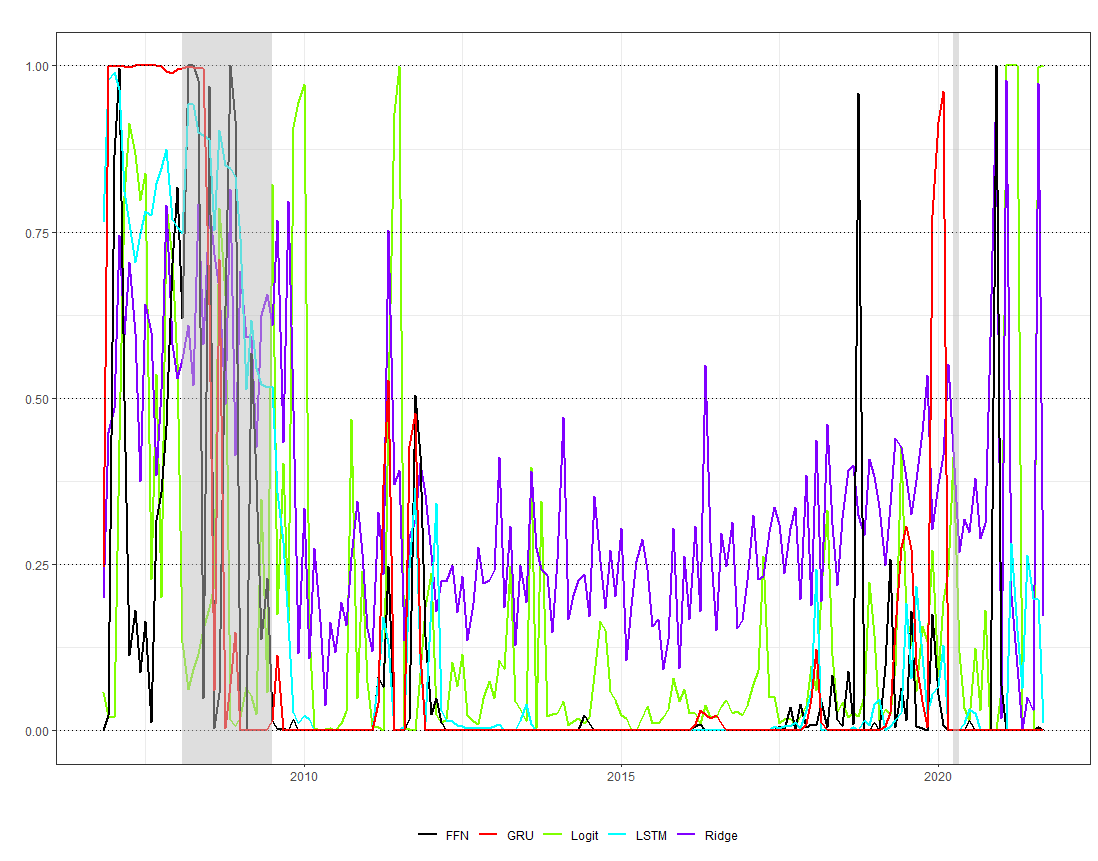} \\
        (E) & \\
        \includegraphics[width=.39\textwidth]{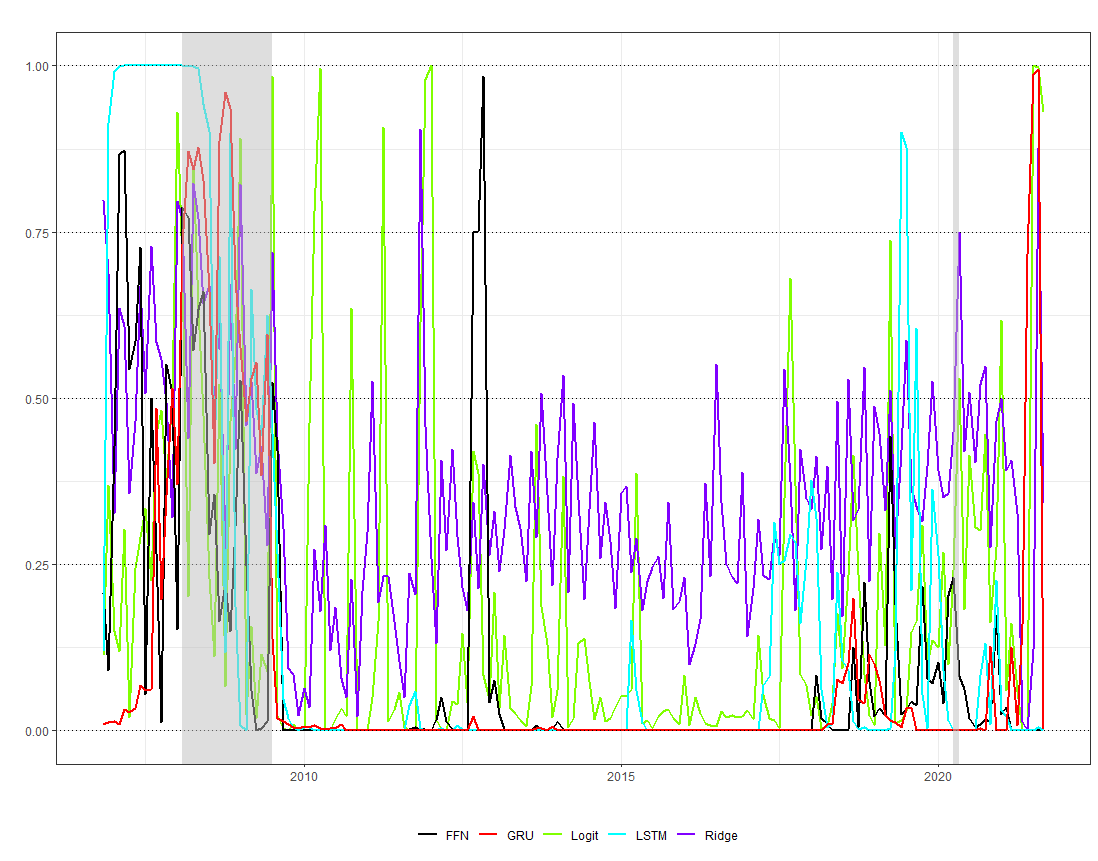} &  \\
    \end{tabular}}
    \caption*{The figure presents out-of-sample recession probabilities for the period from November 2006 to October 2021:: Panel (A) displays the predicted recession probabilities for the nowcasting forecast horizon, comparing the logit model, ridge logit model, FFN, LSTM, and GRU. Panels (B), (C), (D), and (E) show the predicted recession probabilities for the immediate-term (1-months-ahead), short-term (3-months-ahead), medium-term (6-months-ahead), and long-term (12-months-ahead) forecast horizon, respectively. The grey-shaded areas depict NBER recession months.}
\end{figure}

Figure 8 displays the out-of-sample forecasts in probabilities from November 2006 to September 2021, accompanied by the grey-shaded areas representing the Great Recession and the Covid-19 recession. Across all panels, similar patterns emerge: the graphs start at high levels before the Great Recession, either remain high or increase during that period, decrease afterward to a range between 0 and 0.25 for an extended period, and then exhibit significant fluctuations around the Covid-19 recession. The models successfully indicate economic and financial downturns during the Great Recession, but an external event like the COVID-19 recession remains largely unpredictable. Depending on the forecast horizon, once the model learns from the impact of the COVID-19 pandemic on the economy, it struggles to interpret this abrupt change and generates alternating forecasts.

\section{Variable importance}

To better understand the underlying factors driving the results, it is important to identify the main variables that have the most significant impact on the prediction. While linear models can refer to estimated coefficients or marginal effects to measure variable importance, more complex models like neural networks require different approaches. \citet{molnar2020} provides a comprehensive list of methods that enable the interpretation of complex machine-learning models. One such method is Local Interpretable Model-Agnostic Explanations (LIME), which suggests using other interpretable models to approximate the predictions locally. For an instance of interest, LIME constructs a new dataset by altering samples variable-wise, drawing from a normal distribution with sample mean and sample standard deviation, and obtains corresponding predictions from the black box model. This modified dataset is then used to train an interpretable model, such as Lasso or a decision tree, which is weighted by the proximity of the sampled instances to the instance of interest. The interpretable model approximates the black box model's predictions at a local level, even though it may not accurately represent the global approximation. 

Another solution stems from cooperative game theory, particularly the Shapley value introduced by \citet{shapley1953}. The Shapley value method involves assigning payouts to players based on their contribution to the overall payout. This cooperative framework resembles a game where players form coalitions and receive profits based on cooperation. In the context of model predictions and their interpretability, the game pertains to the prediction task, and the payout represents the difference between the actual prediction for that data point and the average prediction across all data points. The players correspond to the variable values of the data point, working together to achieve the payout, which means predicting a specific value. To calculate the Shapley value for a specific variable value, all possible coalitions of variable values, excluding the variable of interest, are formed for each data point. The variable values outside a coalition are substituted with random values of those variables from data to generate a prediction. The predictions are computed for each coalition, both with and without the variable value of interest. The difference between these predictions represents the marginal contribution of the variable value for that coalition. This computation is repeated for all possible coalitions, and the average of the marginal contributions across all coalitions yields the Shapley value for that variable value. Finally, the Shapley values for a variable can be averaged across data points to assess the relative importance of variables compared to each other.

The Shapley value stands out as the sole explanation method supported by a robust theory that satisfies the axioms of efficiency, symmetry, dummy, and additivity to allow for fair distribution of the contributions among the variables (\citet{molnar2020}). Conversely, methods like LIME rely on the assumption of local linear behavior of the machine learning model but need a theoretical justification for why this approach is effective. \citet{lundberg2017} propose an alternative estimation method, the Shapley additive explanations (SHAP), which uses a kernel-based estimation approach for Shapley values (KernelSHAP). SHAP introduces an innovative aspect by presenting the Shapley value explanation as an additive variable attribution method, which can be viewed as a linear model. This perspective establishes a connection between LIME and Shapley values. In this context, SHAP defines the explanation model as follows:

\begin{align*}
    f(c')=\alpha_0+\sum_{j=1}^{K}\alpha_j c'_j
\end{align*}

In the provided equation, the explanation model is denoted as $f$, the coalition vector as $c'\in\{0,1\}^K$, the maximum coalition size as $K$, and the Shapley value for a variable $j$ represented by $\alpha_j$. The coalition vector $c$ indicates the presence or absence of variable values, with a value of 1 representing presence and 0 indicating absence. 

KernelSHAP is a method that estimates the contributions of individual variable values to the predictions by first generating a random coalition $c'\in\{0,1\}^K$ consisting of K members by flipping a coin multiple times until we obtain a sequence of 0's and 1's. The sampled coalition of size $K$ is then used as a data point for the regression model. In this model, the target is the prediction for a coalition, whereby in the case of $c_k=0$, the absent variable value is substituted with random variable values from the data. This process is repeated for each data point. Then Shapley-compliant weights are computed according to the SHAP kernel proposed by \citet{lundberg2017}. Finally, a weighted linear regression model is fitted on the modified data, and the model's estimated coefficients are the Shapley values.

The main distinction from LIME lies in how instances are weighted in the regression model. In LIME, the weighting is determined based on the proximity of instances to the original instance. The closer an instance is, the higher its weight in LIME. On the other hand, SHAP assigns weights to sampled instances based on the weight they would receive in the Shapley value estimation. Small coalitions (with fewer 1's) and large coalitions (with many 1's) receive the highest weights in SHAP. The underlying intuition is that the most knowledge about individual variables can be obtained when their effects can be studied in isolation. For a coalition consisting of a single variable, the isolated main effect of that variable on the prediction can be observed. When a coalition includes all variables except one, it reports the total effect of that particular variable, including the main effect and variable interactions. However, if a coalition comprises half the variables, it provides limited insight into the contribution of an individual variable due to the numerous possible coalitions with half of the variables. 

The SHAP values for logit models are computed using the KernelSHAP method using KernelExplainer from the SHAP package in Python. In contrast, the SHAP values for deep learning models are approximated by an enhanced version of the DeepLIFT algorithm introduced by \citet{shrikumar2017} using DeepExplainer from the same package. DeepLIFT is a method used to analyze the output prediction of a neural network for a specific input. It achieves this by backpropagating the contributions of all units in the network to each input variable. DeepLIFT compares the activation of each unit to its reference activation, where for each variable, its sample mean is used as the reference input, and contribution scores are assigned based on the difference between them. Unlike other approaches, DeepLIFT can uncover dependencies that may be overlooked. It can also consider positive and negative contributions separately. Additionally, the scores can be efficiently computed in a single backward pass.

SHAP method generates a matrix of Shapley values based on the number of data points and variables. These Shapley values can create global explanations by averaging the absolute Shapley values per variable across the twelve-time steps and data points. This process is repeated for the entire out-of-sample period. To compare the SHAP results of neural network models to linear models, GRU and the ridge logit model are selected for the subsequent analysis, as they perform better than the others in their respective groups of models. The results for GRUs are presented in Figure 9, which depicts the distribution of average absolute SHAP values across variables for the entire out-of-sample period. The variables are ranked based on their medians, providing insights into their importance for different forecast horizons.  

The ranking of variables may vary over time, but it offers an initial understanding of the key drivers influencing the prediction results for different forecast horizons. Variables associated with financial conditions such as the S\&P 500 index and term spread, and macroeconomic variables related to GDP, inflation, and the housing market, commonly regarded as important recession indicators, consistently rank high (\citet{estrella1998}). These findings provide valuable insights into the black box and suggest that neural networks effectively capture the patterns inherent in the dynamics of the frequently cited recession indicators.

Figure 10 presents the SHAP results for the ridge logit model. There are three major differences, among others, between the results in Figures 9 and 10. Firstly, the order of variables differs for each forecast horizon between the two types of models. While some variables appear in the leading groups for both models, their specific order varies significantly. For instance, the money supply M2 is influential in the ridge model but less prominent in the GRU model. Spearman's rank correlations between them range from -0.24 to 0.10, indicating a weak correlation.

\begin{landscape}
\begin{figure}[!ht]
\caption{SHAP values of the predictors: GRU}
    \centering
    \renewcommand{\arraystretch}{0.1}
    \begin{tabular}{l l l}
        (A) & (B) & (C)\\
        \includegraphics[width=.35\textwidth]{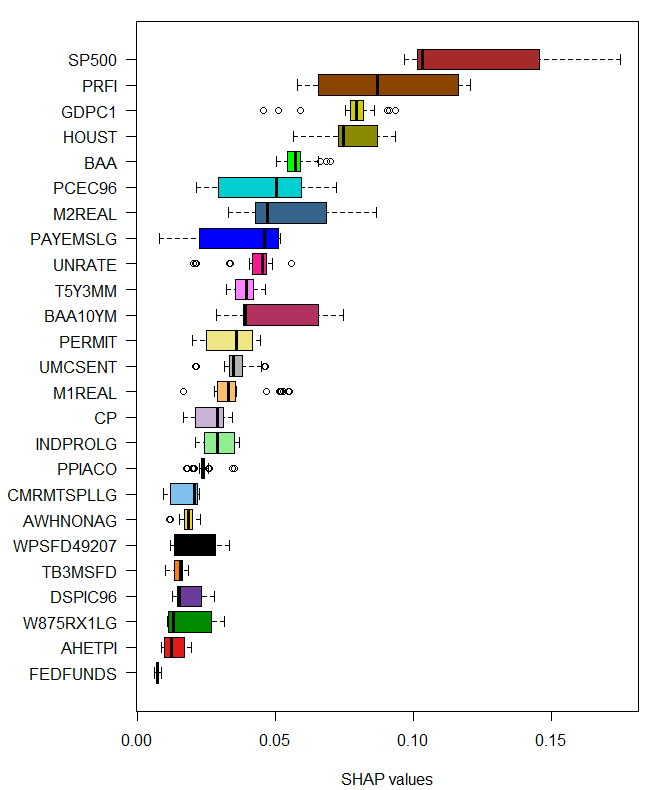} &  \includegraphics[width=.35\textwidth]{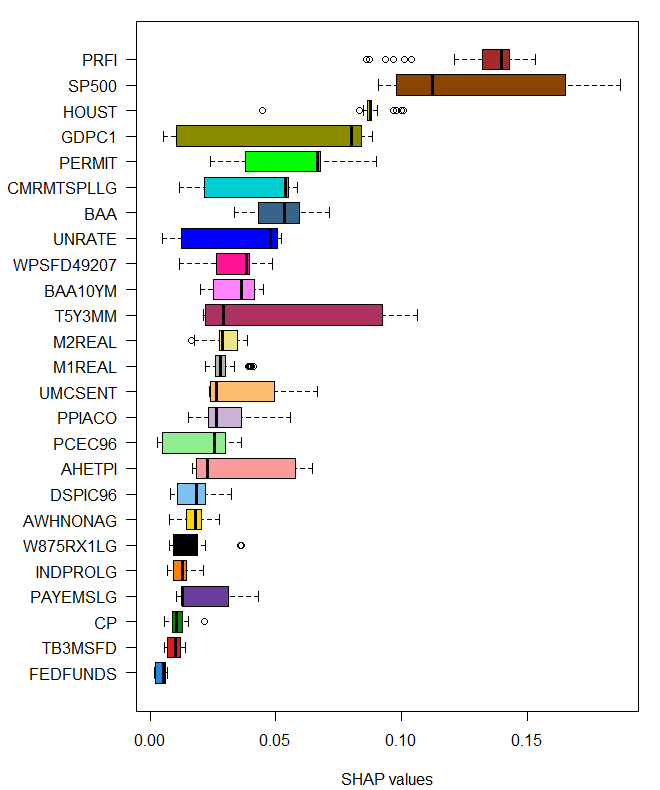} & \includegraphics[width=.35\textwidth]{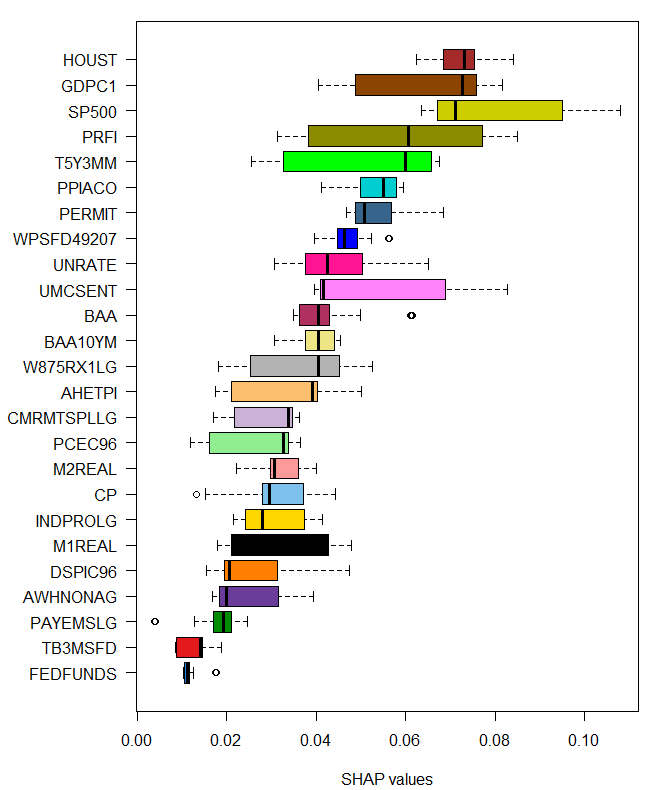} \\
        (D) & (E) & \\
         \includegraphics[width=.35\textwidth]{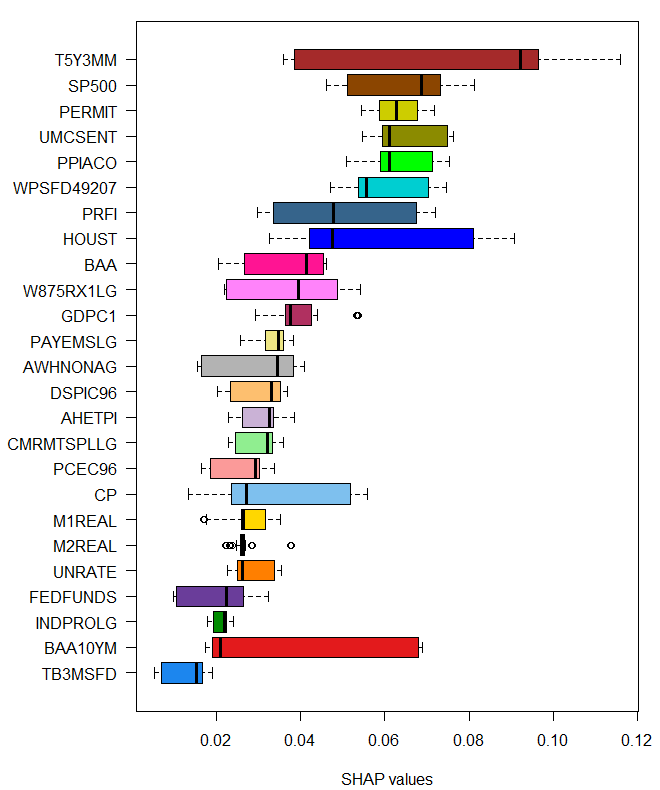} & \includegraphics[width=.35\textwidth]{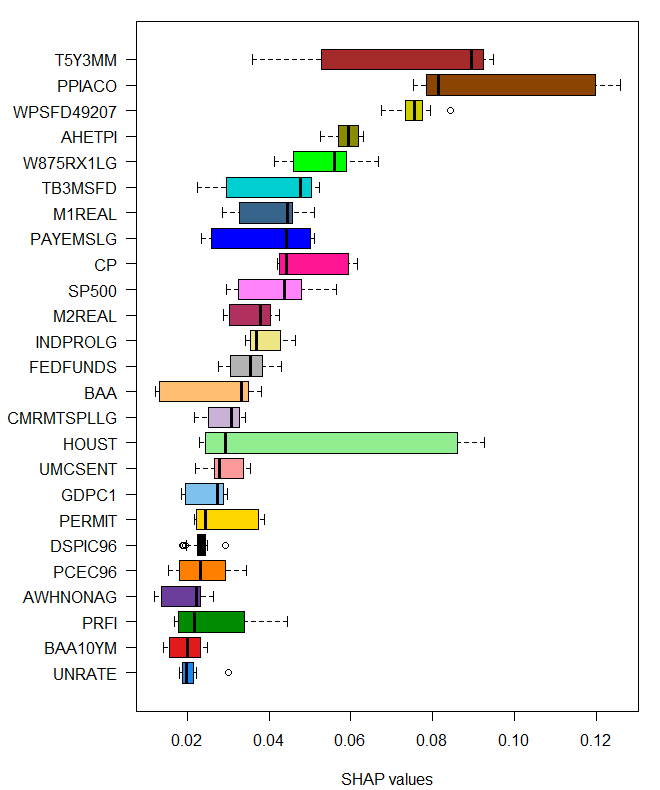} & \\
    \end{tabular}
    \caption*{The figure presents the boxplots of the GRU-based absolute average SHAP values of the predictors and their medians in descending order: Panel (A) to (E) display absolute average SHAP values of the predictors for the nowcasting, immediate-term (1-month-ahead), short-term (3-months-ahead), medium-term (6-months-ahead), and long-term (12-months-ahead) forecast horizon, respectively.}
\end{figure}
\end{landscape}

\begin{landscape}
\begin{figure}[!ht]
    \caption{SHAP values of the predictors: ridge logistic regression}
    \centering
    \renewcommand{\arraystretch}{0.1}
    \begin{tabular}{l l l}
        (A) & (B) & (C)\\
        \includegraphics[width=.35\textwidth]{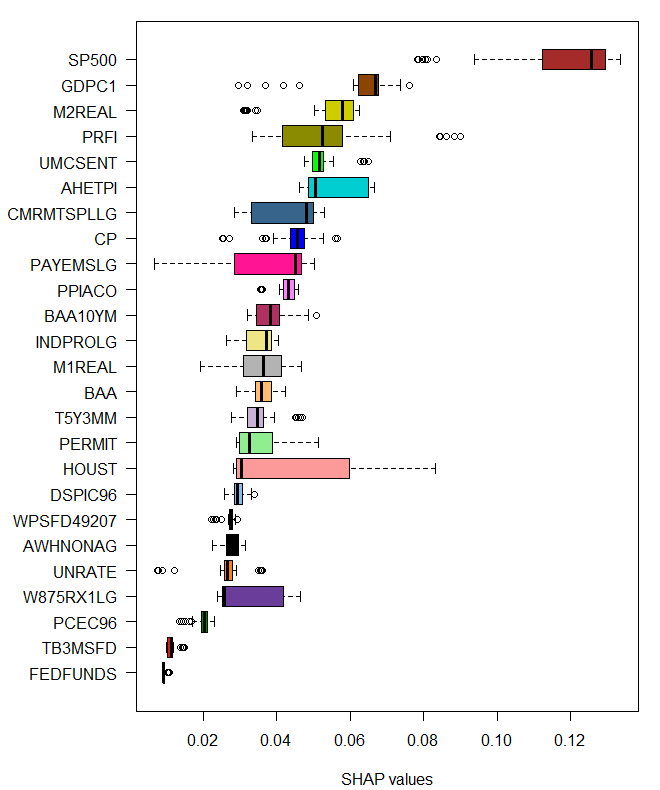} &  \includegraphics[width=.35\textwidth]{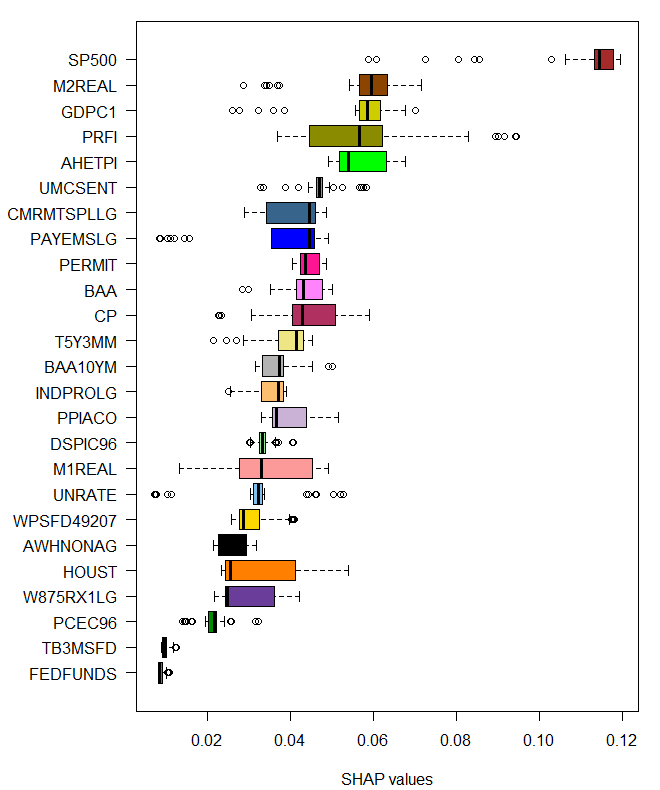} & \includegraphics[width=.35\textwidth]{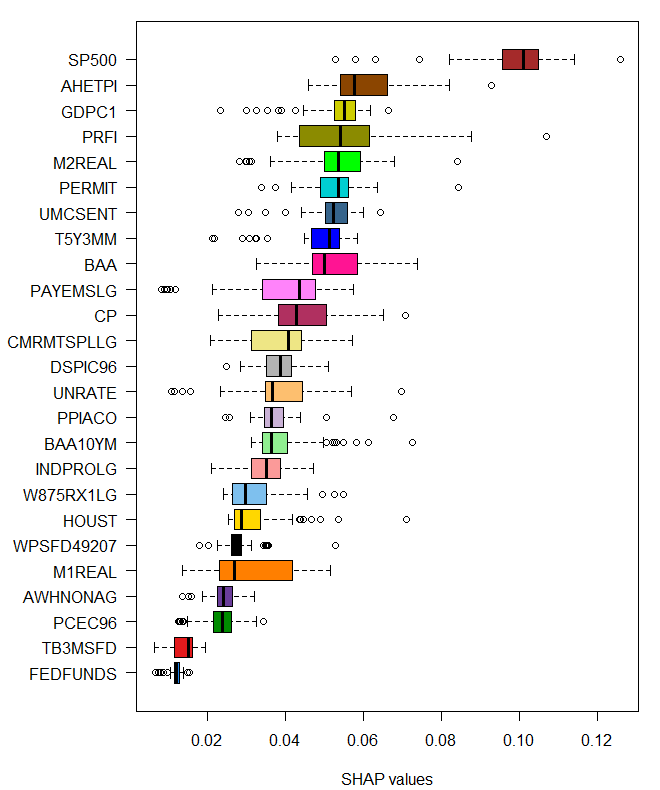} \\
        (D) & (E) & \\
         \includegraphics[width=.35\textwidth]{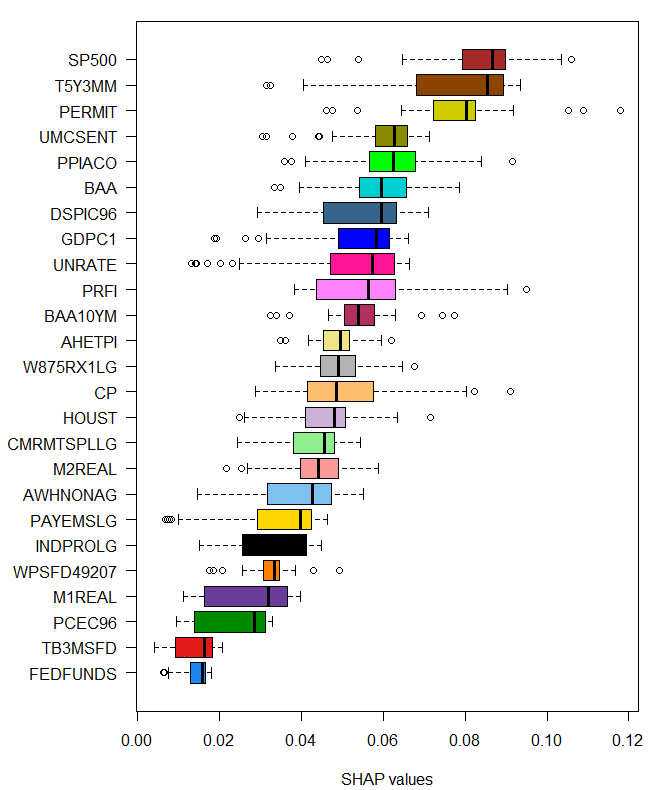} & \includegraphics[width=.35\textwidth]{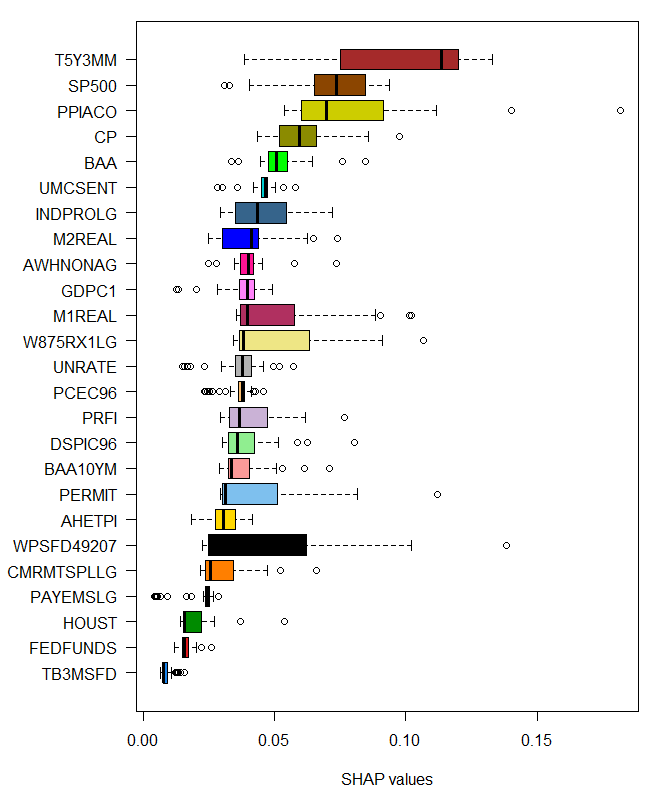} & \\
    \end{tabular}
    \caption*{The figure shows the boxplots of the ridge model-based absolute average SHAP values of the predictors and their medians in descending order: Panel (A) to (E) display absolute average SHAP values of the predictors for the nowcasting, immediate-term (1-month-ahead), short-term (3-months-ahead), medium-term (6-months-ahead), and long-term (12-months-ahead) forecast horizon, respectively}
\end{figure}
\end{landscape}

\noindent This suggests that GRU and ridge models assign different weights to the variables. This difference could be attributed to the neural network's ability to capture nonlinear relationships, allowing for varied weighting of the variables based on different patterns. Secondly, the SHAP values of the GRU models exhibit a more even distribution among the variables than the ridge model. This is reflected in the smaller variation in medians between the variables. The distinction becomes particularly pronounced in the short-term setting (up to 3 months), where the ridge model heavily depends on the S\&P 500 index. Additionally, the ridge model exhibits numerous outliers in the SHAP values, indicating its high sensitivity to changes in macroeconomic and financial conditions in SHAP value estimation. In contrast, this highlights GRU as a more robust modeling framework. Finally, the within variation of the GRU-based SHAP values for some variables is significantly larger than others. Certain variables demonstrate little change in their SHAP values, while others exhibit significant variation. This phenomenon is less prominent in the outcomes of the ridge model. The variances are more uniformly spread, evident in the roughly equal sizes of boxes across variables. This is another indication of the GRU's more resilient modeling framework. The GRU can adapt to shifts in economic conditions, adjusting the weights of specific variables accordingly. In contrast, the ridge model has relatively fixed variable orders with limited flexibility for variation.

The marginal effect analysis is another option to evaluate the variable importance of linear models. Since deep learning models can not be interpreted in terms of marginal effects, the average marginal effects of the ridge model are computed for the out-of-sample period and plotted in Figure 11. The Spearman's rank correlations between the variable series for the ridge model, evaluated using either SHAP values or marginal effects, range from 0.22 to 0.51. The magnitudes of the correlations are relatively small. Still, the previous findings, especially the first one about the leading recession indicators, also apply to the ridge model evaluated by the marginal effects. The averaged LIME values in Figure 12, which may be regarded as the marginal effect version of GRU, reaffirm the main findings discussed above. The only minor distinction is that it emphasizes the term spread as the comprehensive recession indicator, both in short-term and long-term contexts.

\begin{landscape}
\begin{figure}[!ht]
\caption{Marginal effects of the predictors: ridge logistic regression}
    \centering
    \renewcommand{\arraystretch}{0.1}
    \begin{tabular}{l l l}
        (A) & (B) & (C)\\
        \includegraphics[width=.35\textwidth]{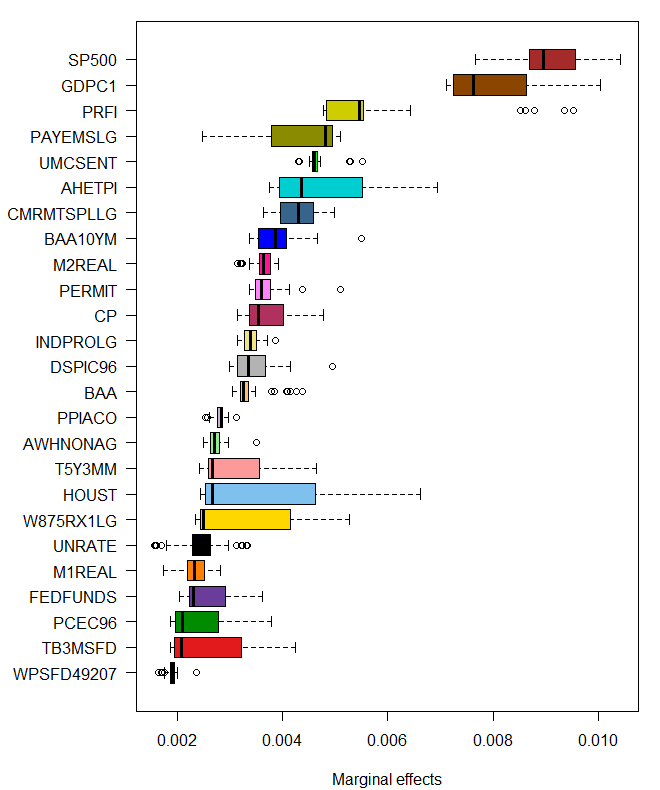} &  \includegraphics[width=.35\textwidth]{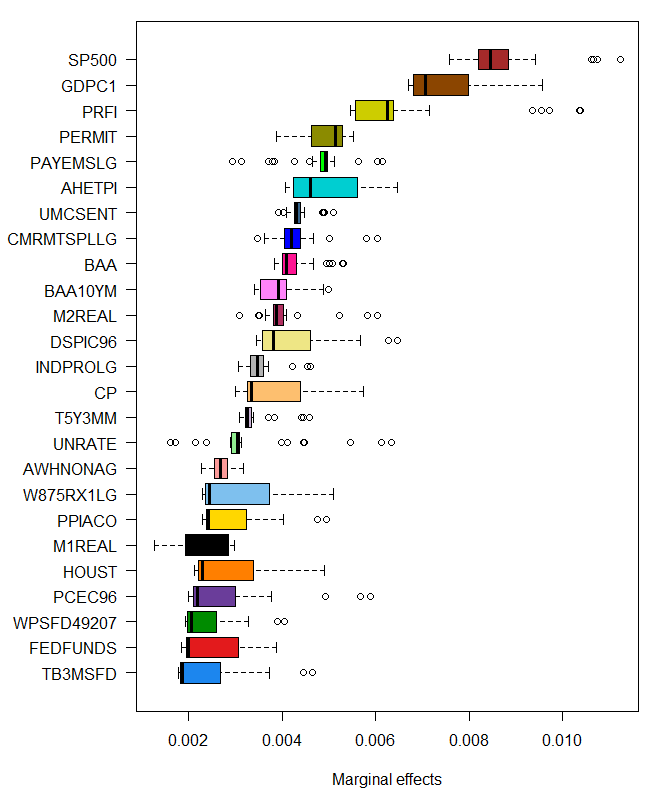} & \includegraphics[width=.35\textwidth]{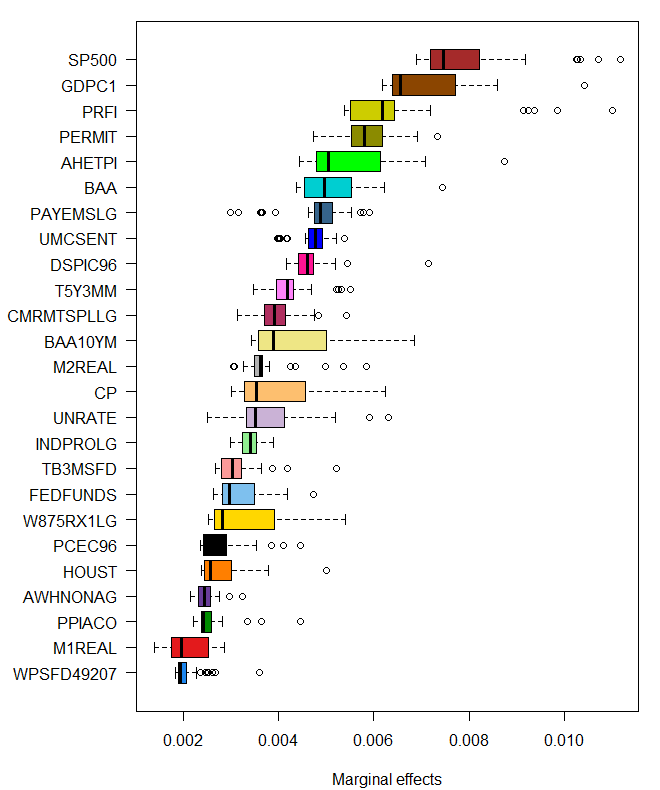} \\
        (D) & (E) & \\
         \includegraphics[width=.35\textwidth]{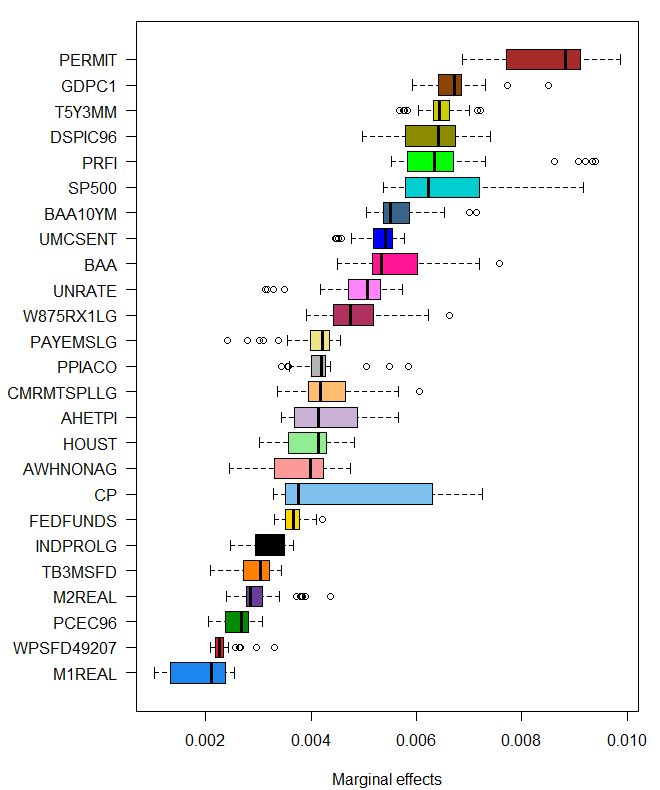} & \includegraphics[width=.35\textwidth]{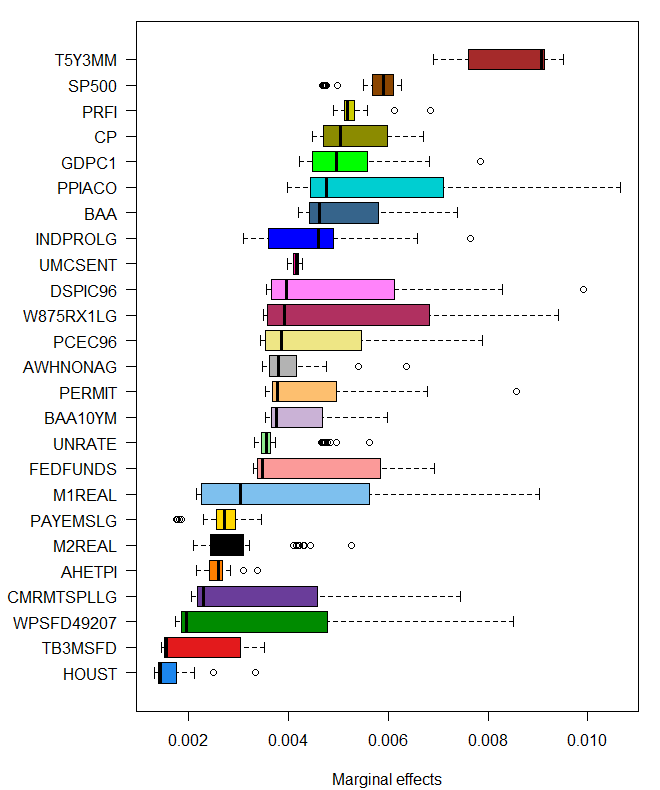} & \\
    \end{tabular}
    \caption*{The figure depicts the boxplots of the ridge model-based marginal effects of the predictors and their medians in descending order: Panel (A) to (E) display marginal effects of the predictors for the nowcasting, immediate-term (1-month-ahead), short-term (3-months-ahead), medium-term (6-months-ahead), and long-term (12-months-ahead) forecast horizon, respectively}
\end{figure}
\end{landscape}

\begin{landscape}
\begin{figure}[!ht]
\caption{LIME values of the predictors: GRU}
    \centering
    \renewcommand{\arraystretch}{0.1}
    \begin{tabular}{l l l}
        (A) & (B) & (C)\\
        \includegraphics[width=.35\textwidth]{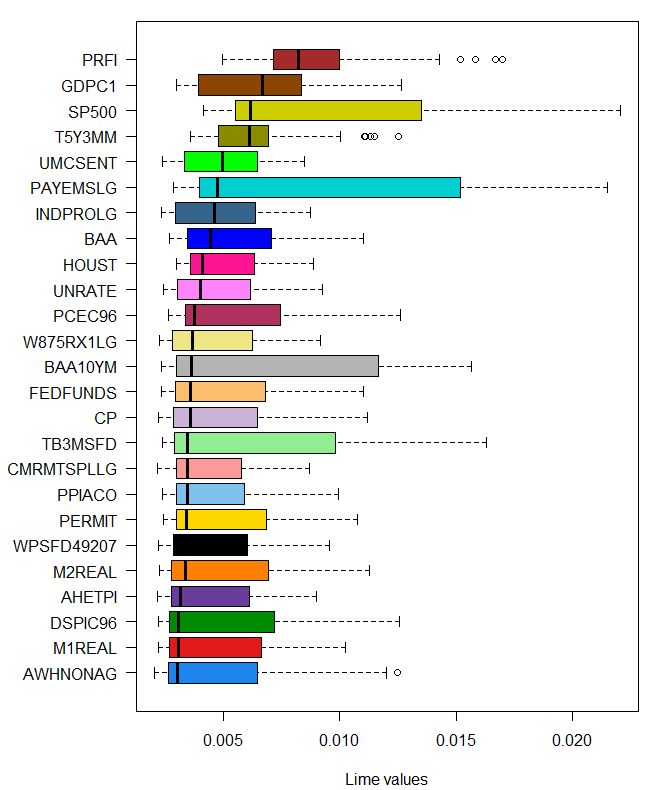} &  \includegraphics[width=.35\textwidth]{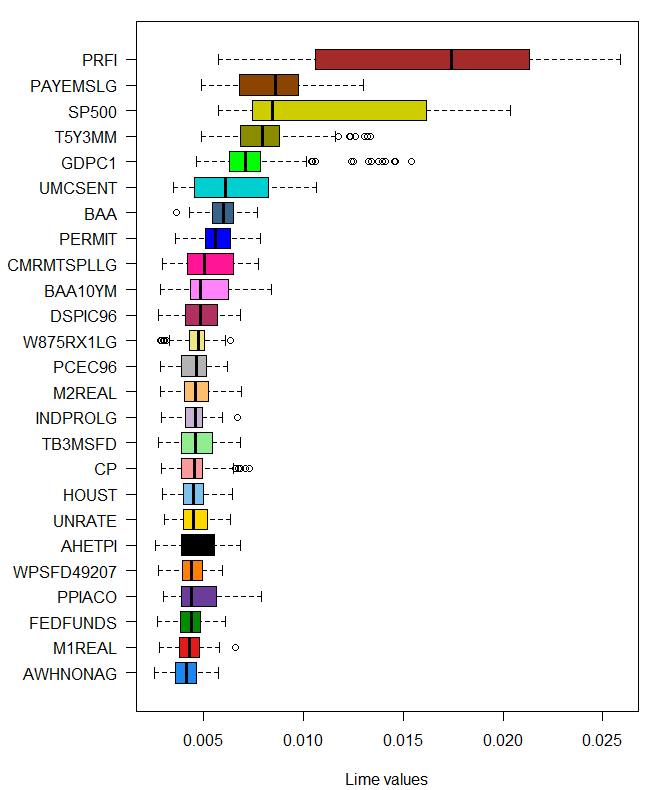} & \includegraphics[width=.35\textwidth]{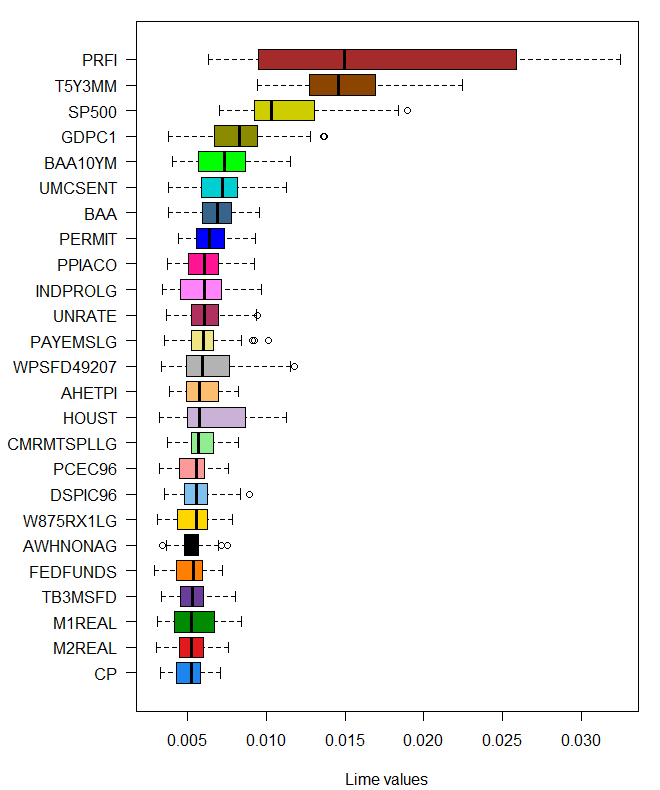} \\
        (D) & (E) & \\
         \includegraphics[width=.35\textwidth]{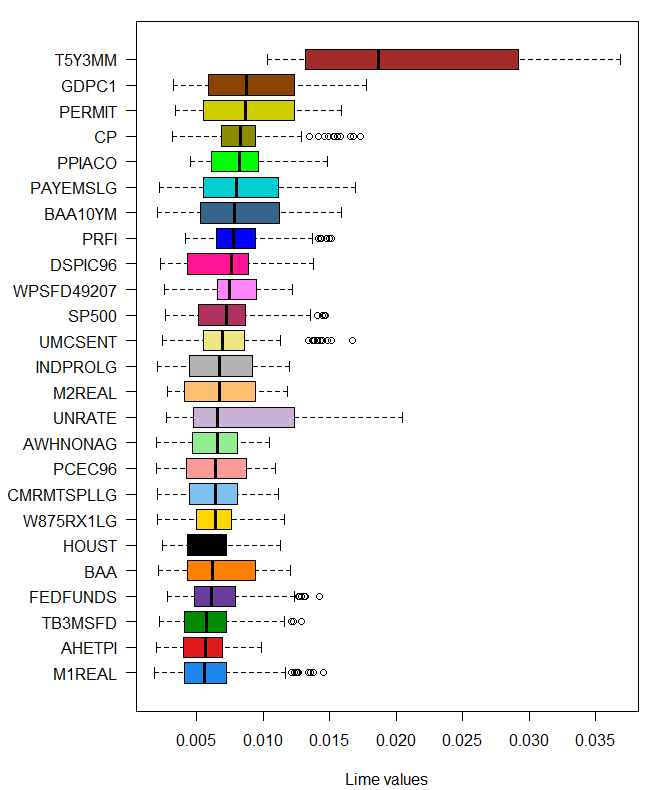} & \includegraphics[width=.35\textwidth]{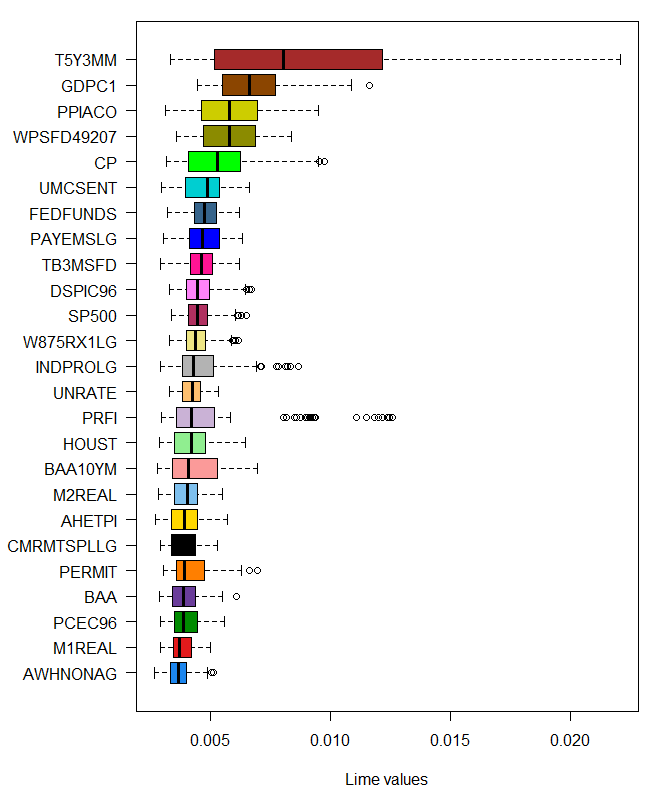} & \\
    \end{tabular}
    \caption*{The figure presents the boxplots of the GRU-based absolute average LIME values of the predictors and their medians in descending order: Panel (A) to (E) display absolute average LIME values of the predictors for the nowcasting, immediate-term (1-month-ahead), short-term (3-months-ahead), medium-term (6-months-ahead), and long-term (12-months-ahead) forecast horizon, respectively.}
\end{figure}
\end{landscape}

\begin{figure}[!ht]
\caption{Dependence plot: Term Spread}
\centering
\includegraphics[width=.75\textwidth]{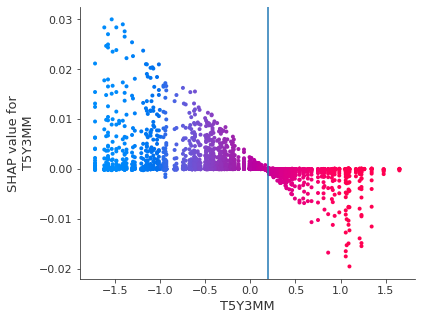}
\caption*{The figure plots the standardized value of the term spread on the x-axis and the corresponding SHAP value on the y-axis. The red and blue colors represent the positive and negative values of the variable, respectively.}
\end{figure}

The SHAP method allows for a detailed analysis of the impact of a single variable on predictions. To anticipate recessions as early as possible, I focus on the long-term forecast horizon (12 months) and consider the term spread as the most relevant and influential variable for GRU to examine its effects on recession predictions for the GRU model. Figure 13 presents the dependence plot of the term spread using the latest available data from October 2021. The x-axis represents the standardized value of the variable (term spread), while the y-axis represents the corresponding SHAP value. Notably, there is a cutoff point around 0.2, above which a higher value of the term spread has a negative impact on the predicted recession probability. In contrast, a value below the cutoff positively influences the predicted probability. Additionally, the farther the value deviates from the cutoff, the stronger the effect of the term spread on the predicted probability becomes. This corroborates the traditional interpretation of the term spread about the recession probability according to \citet{estrella1996} and indicates that advanced neural network models like GRU uncover and utilize this relationship.

\section{Conclusion}

This research examines the real-time predictability of neural network models, compared to linear models, for the recent two US recessions, the Great Recession and the COVID-19 recession. Three different neural network models are trained and updated throughout the study: a standard feed-forward neural network model (FFN), as well as two types of recurrent neural networks, LSTM and GRU, designed to capture temporal dependencies in time series data effectively. The performance of these models is evaluated using out-of-sample forecasts and compared to standard and ridge logit models. Additionally, the SHAP method ranks the predictors based on their importance for each forecast horizon, providing initial insights into the most influential predictors. The results are then compared to SHAP results obtained from the ridge logit model. The main findings are validated using the LIME and marginal effect methods. For in-depth analysis, the term spread, the most influential variable for the long-term forecast horizon, is chosen to investigate the impact of the variable on the recession probability. 

This paper introduces two main contributions. Firstly, it focuses on LSTM and GRU, specialized recurrent neural network models that address issues like exploding and vanishing gradients in standard recurrent neural networks. The performance of these models is compared to FFN and the logit variants in the context of recession forecasting. Secondly, the paper employs the SHAP method to assess the variable importance in GRU and the ridge logit model for different forecast horizons. Given neural networks' commonly perceived black-box nature, the SHAP method delivers a theoretically sound framework to provide insights into the underlying rationales influencing the predictions. 

The three main findings are as follows: Firstly, LSTM and GRU demonstrate strong out-of-sample performance in recession forecasting across five forecast horizons, particularly in long-term predictions. Secondly, there are differences in how GRU and the ridge logit model evaluate the variable importance. The variable order differs between GRU and the ridge model for each forecast horizon. While some variables are significant in both models, their ranking varies notably. This suggests that GRU and the ridge model assign different weights to variables, potentially due to the neural network's capacity to capture nonlinear relationships and assign varied weights based on distinct patterns. Furthermore, the SHAP values of GRU models show a more balanced distribution among variables than the ridge model. This is evident in the smaller variation in medians between variables. The difference is particularly noticeable in the short-term scenario (up to 3 months), where the ridge model heavily relies on the S\&P 500 index. Moreover, the ridge model displays numerous outliers in SHAP values, indicating high sensitivity to changes in macroeconomic and financial conditions. In contrast, this emphasizes GRU's robustness. The within variation of GRU-based SHAP values for some variables is significantly larger than for others. Certain variables show minimal change in SHAP values, while others exhibit considerable variation. This contrast is less pronounced in the ridge model outcomes, where variances are more evenly distributed, reflected in similar box sizes across variables. This suggests that the GRU has a more adaptable modeling framework capable of adjusting variable weights in response to shifts in economic conditions. Conversely, the ridge model has relatively fixed variable orders with limited flexibility for variation. Lastly, although the primary predictors for GRU and ridge logit models show slight differences, key indicators such as the S\&P 500 index, real GDP, and private residential fixed investment consistently play a significant role in short-term predictions (up to 3 months). For longer-term forecasts (6 months or more), the term spread and producer price index become more prominent. Other interpretation methods support these results, such as the local interpretable model-agnostic explanations (LIME) for GRU and the marginal effects for the ridge logit model. 

\vspace{5cm}

\noindent \textbf{Declaration of generative AI and AI-assisted technologies in the writing process} \\
During the preparation of this work the author used Grammarly in order to improve language and readability. After using this tool, the author reviewed and edited the content as needed and takes full responsibility for the content of the publication. 

\clearpage
\bibliographystyle{IBB_IJF} 
\bibliography{IBB_IJF}





\end{document}